\documentclass[journal=mamobx,manuscript=article]{achemso}

\usepackage[version=3]{mhchem} 
\usepackage{siunitx}
\usepackage{xcolor}

\author{Justin Tauber}
\affiliation[Wageningen University]
{Physical Chemistry and Soft Matter, Wageningen University and Research, Stippeneng 4, 6708 WE Wageningen, the Netherlands}
\author{Lorenzo Rovigatti}
\affiliation[Sapienza]
{Dipartimento di Fisica, Sapienza-Universit\`{a} di Roma, Piazzale A. Moro 2, 00185 Roma, Italy}
\author{Simone Dussi}
\affiliation[Wageningen University]
{Physical Chemistry and Soft Matter, Wageningen University and Research, Stippeneng 4, 6708 WE Wageningen, the Netherlands}
\alsoaffiliation[Harvard]
{John A. Paulson School of Engineering and Applied Sciences, Harvard University, Cambridge, MA 02138, USA}
\author{Jasper van der Gucht}
\email{jasper.vandergucht@wur.nl}
\affiliation[Wageningen University]
{Physical Chemistry and Soft Matter, Wageningen University and Research, Stippeneng 4, 6708 WE Wageningen, the Netherlands}

\title[DNSwelling]
  {Sharing the load: stress redistribution governs fracture of polymer double networks}

\abbreviations{IR,NMR,UV}

\begin{document}


\begin{abstract}
The stress response of polymer double networks depends not only on the properties of the constituent networks, but also on the interactions arising between them. Here we demonstrate, via coarse-grained simulations, that both their global stress response and their microscopic fracture mechanics are governed by load sharing through these inter-network interactions. By comparing our results with affine predictions, where stress redistribution is by definition homogeneous, we show that stress redistribution is highly inhomogeneous. In particular, the affine prediction overestimates the fraction of broken chains by almost an order of magnitude. Furthermore, homogeneous stress distribution predicts a single fracture process, while in our simulations fracture of sacrificial chains takes place in two steps governed by load sharing within a network and between networks, respectively. Our results thus provide a detailed microscopic picture of how inhomogeneous stress redistribution after rupture of chains governs the fracture of polymer double networks.
\end{abstract}


\section{Introduction}

By consecutively cross-linking two interpenetrating polymer networks, a composite material is created which is commonly referred to as a (polymer) double network (DN).~\cite{Gong2003,Ducrot2014} In many DNs, the two underlying networks do not share any inter-network crosslinkers and are only topologically constrained at the chain-level~\cite{Gong2003,Nakajima2012}. For this reason, DNs can be considered as a molecular composite~\cite{Millereau2018}. DNs have attracted considerable interest due to the significant enhancement in their (linear) stiffness, strength, and fracture toughness compared to single networks (SNs)~\cite{Gong2003, Tanaka2007, Brown2007,Xin2013,Ahmed2014,Millereau2018}. For example, through this procedure hydrogels can be constructed that have a mechanical response similar to that of an elastomer~\cite{Gong2010}. 

Experiments reveal that stiff, strong and tough DNs are created when the first network, or sacrificial network, is stiff and weak, while the second network, or matrix network, is soft and stretchable.~\cite{Gong2010} To make networks with these properties, one can vary the type and concentration of monomers and crosslinkers in both networks~\cite{Ahmed2014}. Additionally, these properties can be controlled by swelling the sacrificial network either by introducing a molecular stent~\cite{Nakajima2012,Matsuda2016} or using the monomer of the second network~\cite{Ducrot2014}. Experiments on a range of systems, varying from elastomers~\cite{Ducrot2014,Millereau2018} to macroscopic networks~\cite{King2019}, suggest that the mechanism through which the enhancement occurs is surprisingly general: sharing of load between the two networks via their topological constraints~\cite{Nakajima2017}.

The corresponding microscopic picture is that, due to the presence of the matrix chains, the expansion of a (microscopic) crack in a DN requires considerably more energy than in an SN.~\cite{Brown2007,Tanaka2007,Xin2013,Nakajima2013a,Ducrot2014} As a consequence, fracture of sacrificial chains in DNs is less likely to lead to the formation of macroscopic cracks and thus global failure.~\cite{Tauber2020a} Instead, the load is transferred (partially) from the sacrificial network to the matrix network surrounding the broken sacrificial polymer chain~\cite{Millereau2018}. Thus, in a DN more sacrificial chains can break prior to global failure compared to an SN. As the intact sacrificial chains in these DNs can still resist deformation, the work required for global failure of the material is increased significantly compared to the individual networks. This concept has been termed the sacrificial bond principle~\cite{Nakajima2017} and is widely accepted as the main microscopic cause for the enhanced mechanical properties in the fracture regime. This microscopic picture for accumulation of damage is confirmed by experiments~\cite{Webber2007,Nakajima2013a,Ducrot2014,Mai2018} and simulations~\cite{Higuchi2018}. However, despite state-of-the art experimental techniques enabling the visualization of stress~\cite{Chen2020}, strain~\cite{Fukao2020,Ducrot2015,VanDerKooij2018} and the accumulation of damage at the local level~\cite{Ducrot2014,Millereau2018,Matsuda2020}, a thorough understanding of how the microscopic processes affect the global material response is still lacking. 

Several constitutive models have been put forward that provide a connection between the evolution of damage and the global stress response of a DN.~\cite{Zhao2012,Liu2016,Vernerey2018,Lavoie2019,Morovati2020} These models can be fitted to experimental data and are also used in the interpretation of the output of mechanophores, i.e. molecular probes that report on the rupture of bonds locally~\cite{Chen2020}. Some of these models~\cite{Vernerey2018,Lavoie2019,Morovati2020}, referred to as statistical damage mechanics models, predict the global response from the evolution of chain-stretch with respect to an initial stretch distribution, assuming affine deformation and breaking of over-stretched chains. As a result, the global mechanical response of a DN predicted by these models is the sum of the response of two (or more) independent and affinely deforming networks. Effectively, these models assume that (statistically or on average) the intra-network load distribution follows the global deformation and that inter-network load redistribution is negligible at the global level. However, in the case of DN mechanics these assumptions deserve some scrutiny, because at first glance they seem to be incompatible with the proposed DN toughening mechanism where inter-network load sharing plays an essential role. In this work we investigate to what extent the microscopic process of redistribution of load, both within and between networks, affects the global mechanical response.

To this end we perform coarse grained simulations of polymer networks, where load redistribution is intrinsically captured. To generate the DNs we replicate a swelling procedure \textit{in silico} which is commonly used to make both hydrogel and elastomer DNs in experiment.~\cite{Nakajima2012,Matsuda2016,Ducrot2014} By deforming the networks we obtain information on both the global stress response and the local stress, the local strain and the accumulation of damage. We show that the \textit{in silico} networks behave in accordance with their experimental counterparts. Subsequently, we compare these simulation results with the affine predictions for the global stress and local damage response. From this comparison we find that in our simulations the microscopic mechanism for damage accumulation differs significantly from the affine approximation, with the affine prediction overestimating the fraction of broken chains by almost an order of magnitude. Furthermore, we show that the accumulation of damage occurs in two steps, one controlled by load sharing within the sacrificial network and one by load sharing between the two networks. Finally, we show that in our simulations load sharing causes an enhanced global mechanical response, in contrast to the affine prediction. We conclude with a discussion of the implications of our findings for the microscopic picture of fracture in DNs and polymer networks in general.

\section{Results and discussion}

\subsection{\textit{In silico} preparation of DNs}

\begin{figure}
    \centering
    \includegraphics{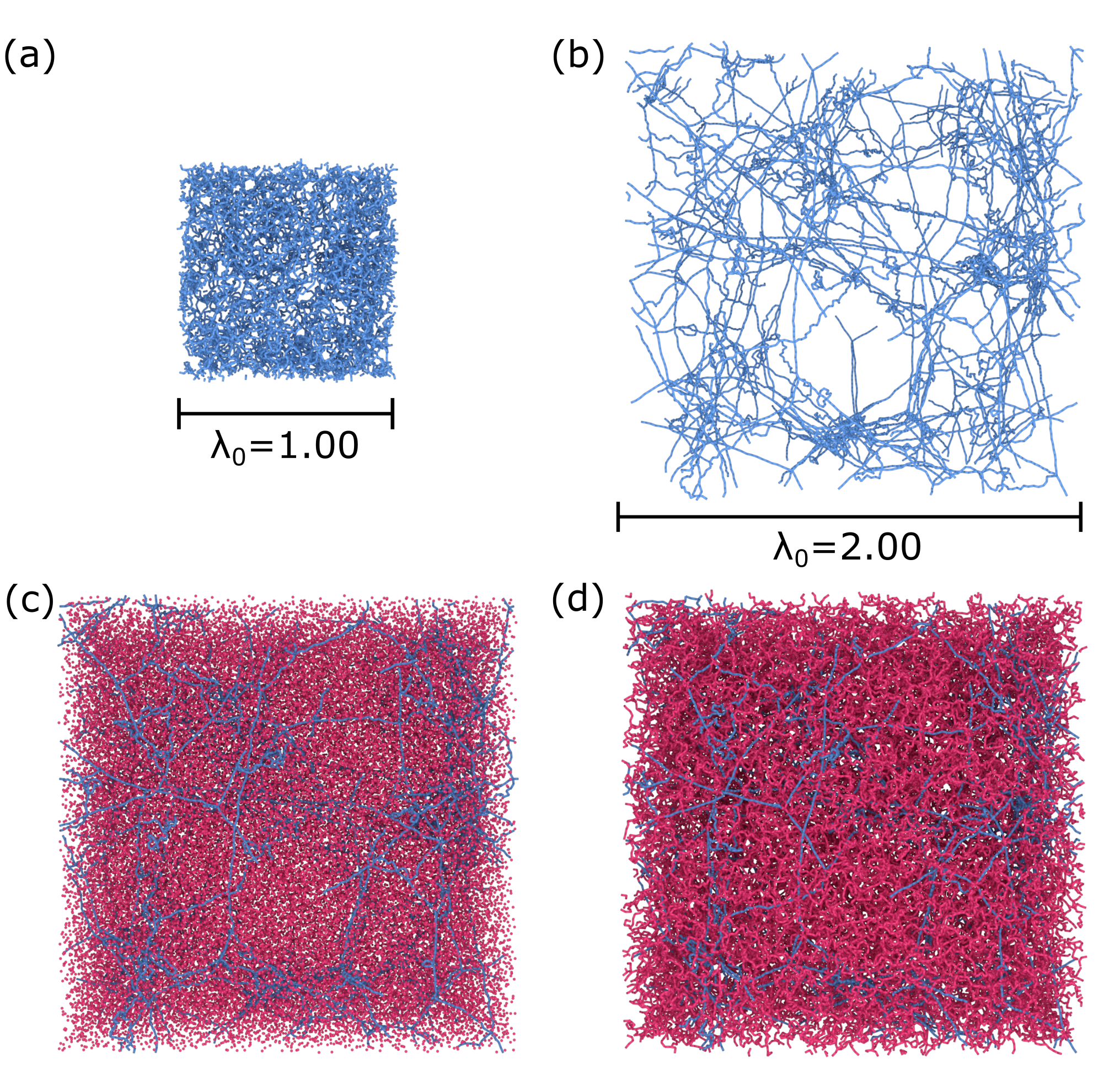}
    \caption{\textit{In silico} double network generation. The sacrificial network is swollen from (a) $\lambda_0=1.00$ to (b) $\lambda=\lambda_0$. In this example $\lambda_0=2.00$. (c) Monomers for the matrix are added at random positions in the swollen sacrificial network such that the number density of the entire system equals $\rho$. (d) The matrix is formed within the sacrificial network with the same crosslinking procedure used for the sacrificial network, but with a lower crosslinker concentration.}
    \label{fig:fig1}
\end{figure}

We prepared our networks following the procedure of Refs.~\cite{Gnan2017,Rovigatti2018,Sorichetti2021}. In particular, the first network, or sacrificial network, is generated from \num{10000} particles of diameter $\sigma$ with a number density $\rho=0.15$ (Fig.~\ref{fig:fig1}(a)). The majority of these particles are bifunctional and can only form linear chains. A fraction $c_1=\SI{5}{\%}$ of the particles are tetrafunctional and can crosslink polymer chains. After network generation, we swell the sacrificial network isotropically up to a swelling ratio $\lambda_0=L_{\text{box}}/L_{0,\text{box}}$ (Fig.~\ref{fig:fig1}(b)), and we add particles for the second network, or matrix (Fig.~\ref{fig:fig1}(c)). The matrix is formed with a crosslinker fraction $c_2=\SI{1}{\percent}$ (Fig.~\ref{fig:fig1}(d)). Because $c_2<c_1$ the sacrificial polymer chains are shorter ($\langle L \rangle=10.9\sigma$) than the matrix chains ($\langle L \rangle=47.3\sigma$) in line with the empirical design rules for creating tough DNs~\cite{Gong2010}. The distribution in chain-lengths is exponential, as expected for a random polymerization process (see Supporting Information for details). A detailed description of the procedure can be found in the method section. We have chosen the parameters based on a trade-off between experimental reality and feasibility of the simulations (see Supporting Information for details).

Note that our protocol for creating DNs differs from other simulation works in several ways. Typically, polymer networks are formed by crosslinking pre-formed chains of given length~\cite{Jang2007,Higuchi2018,Li2020,Yin2020}, rather than using a random-polymerization-like procedure as we do. Furthermore, we prepare DNs via sequential polymerizations, that conceptually resemble the experimental protocols~\cite{Nakajima2012,Ducrot2014}, instead of the simultaneous assembly of both networks as previously done~\cite{Jang2007,Higuchi2018,Li2020}. There are other examples of \textit{in silico} DNs generated by swelling. However, differently from our procedure, either inter-network crosslinking is allowed~\cite{Wang2017}, or only the bonds of crosslinkers are allowed to break~\cite{Yin2020}.

\subsection{Mechanical response of \textit{in silico} DNs}

\begin{figure}
    \centering
    \includegraphics{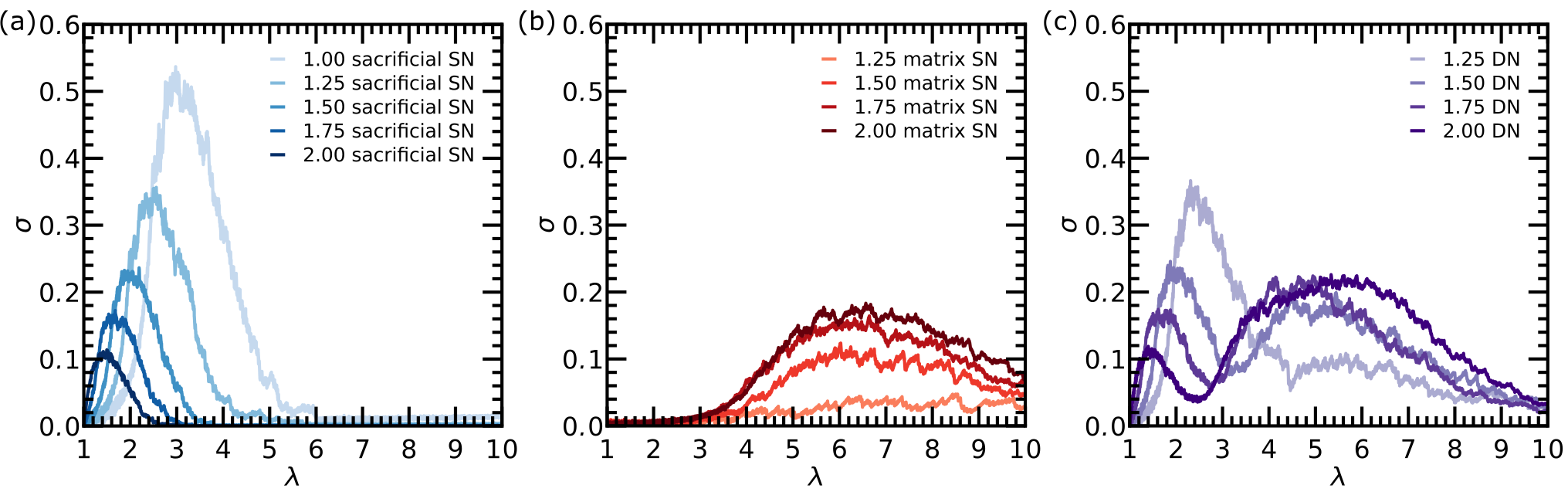}
    \caption{Mechanical response of SNs and DNs for a range of swelling ratios (see legends). We plot the engineering stress $\sigma$ versus the global stretch $\lambda$ for (a) sacrificial SNs ($c_1=5\%$), (b) Matrix SNs ($c_2=1\%$), and (c) DNs ($c_1=5\%$, $c_2=1\%$). Note that the matrix SNs are generated by removing the sacrificial chains from the DNs.}
    \label{fig:fig2}
\end{figure}

To obtain the mechanical response of our networks, we perform a uniaxial extension at a constant strain-rate and a constant volume i.e. we impose a Poisson ratio $\nu=0.5$. We do this for both the DNs and the stand-alone networks. To facilitate comparison with experimental work, we plot the engineering stress $\sigma$ which is calculated by dividing the deviatoric (true) stress by the global stretch $\lambda$ (see method section for details). All results are reported in reduced (Lennard-Jones) units.

Following one of the curves in Fig.~\ref{fig:fig2}(a) (e.g. $\lambda_0=1.00$) we can identify four mechanical regimes. After a short linear response at low strain (the linear elastic regime), the network becomes strain-stiffening, as is expected for entropic springs, around $\lambda=1.50$ (the non-linear elastic regime). Subsequently, strain-softening starts from $\lambda=2.25$, induced by breaking of chains, until the maximum strength $\sigma_{\text{max}}$ of the sacrificial SN is reached at $\lambda=3.35$ (the strain-softening regime). After this point, the stress drops rapidly, indicating that significant damage is done to the network, cracks start to propagate and the capability to carry load is lost (the crack propagation regime). Swelling the networks (without adding the matrix), increases the linear modulus (see Supporting Information) and decreases the onset of strain-stiffening, the stretch at maximum strength and the maximum strength. All these effects can be explained by the fact that network swelling leads to pre-stretching of the polymer chains, so that polymers are tensed already before applying uniaxial deformation, so that less additional stress is needed to induce strain-stiffening and rupture. This has been observed also in experiment.~\cite{Matsuda2016,Millereau2018} Note that at the largest swelling ratio $\lambda_0=2.00$ a few sacrificial chains break already during the swelling procedure ($\sim\SI{0.5}{\percent}$ of all chains).

For the stand-alone matrix networks (matrix SN), obtained by removing the sacrificial network, we find the same mechanical regimes as for the sacrificial SN (Fig.~\ref{fig:fig2}(b)). However, as on average the matrix chains are longer than the sacrificial chains, the onset of strain stiffening and the maximum stress are found at higher strains. Because the matrix networks are formed after the swelling procedure (and thus carry no significant pre-stretch), we do not find a significant shift in the onset of strain stiffening or the strain at maximum strength with $\lambda_0$. We do find a dependency of $\sigma_{\text{max}}$ on $\lambda_0$, which is caused by the increase in the matrix monomer density $\rho_2$ with the swelling ratio: $\rho_2 = \rho - \rho_{1} = \rho (1 - 1/\lambda_0^3)$. In other words, the polymer chain density in the matrix increases with $\lambda_0$, providing more chains to resist elongation. 

The mechanical response of the DNs (Fig.~\ref{fig:fig2}(c)) is clearly influenced by both the sacrificial network, which dominates at low strain, and the matrix network, which dominates at high strain. The loop in stress at intermediate strains marks the transition between these two regimes. A similar transition is observed for some hydrogels and elastomers in experiments in the form of a plateau with a constant stress after a certain ``yield'' point.~\cite{Matsuda2016,Millereau2018} Such an extended fracture response is atypical for hydrogels and elastomers, which normally fracture in an abrupt manner after reaching their maximum strength, i.e. brittle fracture. In analogy to the fracture response of various metals, this extended fracture response is referred to as ductile. In polymer double networks this plateau is caused by the separation of the material into a soft and highly stretched region, in which many sacrificial bonds are broken, and a stiff and weakly stretched region, in which the sacrificial network is still intact, also referred to as necking. A force balance between these two regions causes the stress to be constant. We do not observe a plateau in our simulations, because our networks are too small to get a separation into a soft (weakly stretched) and a stiff (highly stretched) region.
In our simulations the maximum strength of the material $\sigma_{\text{max}}$ is determined by the stress peak either before or after the loop, depending on $\lambda_0$. We think this is indicative of the transition from brittle to ductile fracture which is observed in experiment as function of $\lambda_0$.~\cite{Millereau2018} For small $\lambda_0$, the first peak is highest, so that in an experimental setting the material will fracture abruptly once the local strain in any part of the network exceeds the strain at $\sigma_\text{max}$. However, for higher $\lambda_0$ when the second peak becomes higher than the first, a coexistence between regions of different (local) $\lambda$ becomes possible at a stress equal to the peak stress of the first network, resulting in a ductile fracture response. The brittle-to-ductile transition would then occur at the point where both peaks are of the same height.

\begin{figure}
    \centering
    \includegraphics{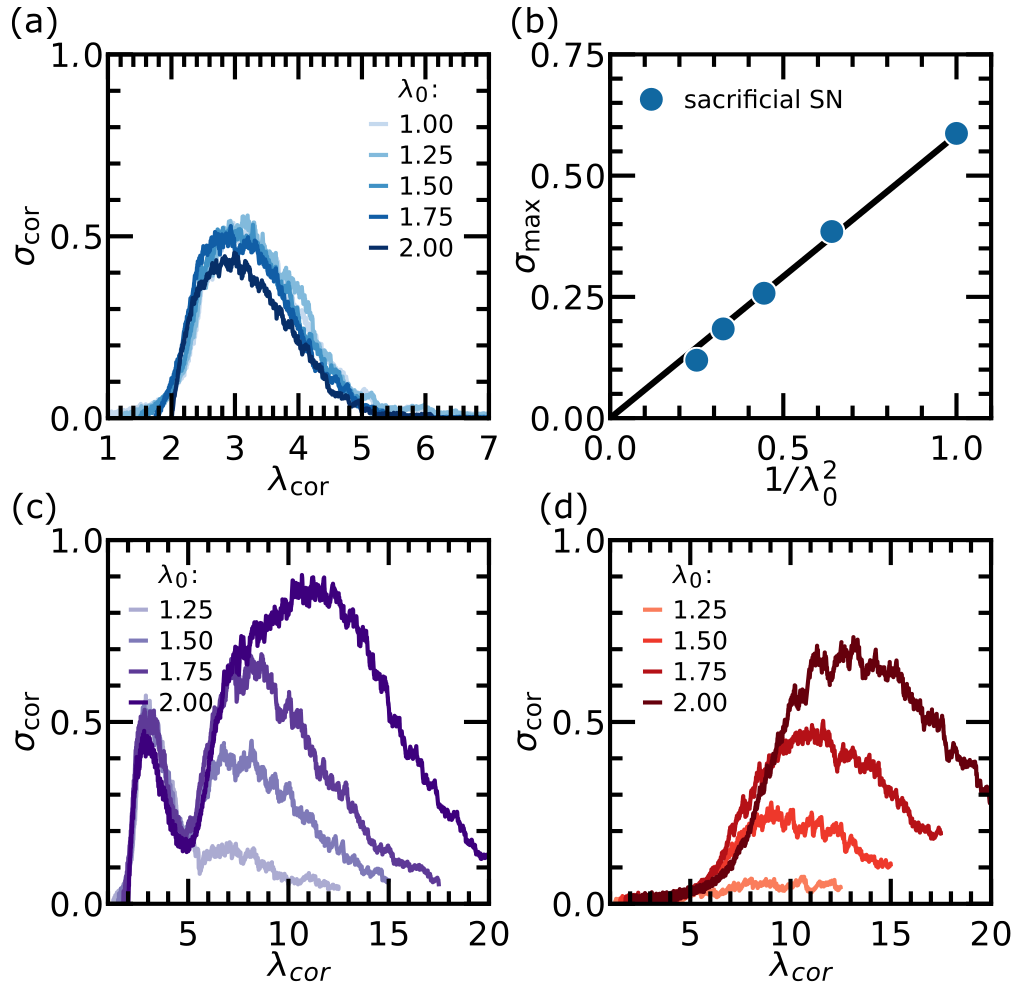}
    \caption{The initial stress response is controlled by the sacrificial network. (a) Rescaled stress-strain curves of sacrificial SNs ($c_1=5\%$), with $\sigma_{\text{cor}}= \sigma\lambda_0^2$ and $\lambda_{\text{cor}}=\lambda\lambda_0$. (b) Maximum stress of sacrificial SN versus $1/\lambda_0^2$, a proxy for strand density. (c) Rescaled stress-strain curves of DNs ($c_1=5\%$, $c_2=1\%$). (d) Rescaled stress-strain curves of the matrix SNs ($c_2=1\%$). }
    \label{fig:fig3}
\end{figure}

We already mentioned that the initial mechanical response of our DNs is dominated by the sacrificial network. Experimental work on hydrogels and swollen elastomers show similar results~\cite{Millereau2018,Nakajima2020} and reveal that the experimental data can be rescaled onto a single master curve based on the areal strand density of the sacrificial network.~\cite{Matsuda2016,Millereau2018} Fig.~\ref{fig:fig3} reveals that also our simulation data can be collapsed on a master curve for both sacrificial SNs and DNs, confirming that at least up to the first peak in stress the response is dominated by the sacrificial network. The rescaling corrects for the increase in pre-stretch ($\lambda_{cor}=\lambda\lambda_0$) and the reduction in areal strand density in the sacrificial network ($\sigma_{cor} = \sigma\lambda_0^2$).
The onset of strain-stiffening in our rescaled curves occurs around $\lambda_{\text{cor}}\approx1.5$ and the peak stress falls around $\lambda_{\text{cor}}\approx3.0$. The latter value is close to the maximum extension limit of our chains under ideal conditions ($\lambda_{\text{limit}}=\sqrt{N}=3.37$ with $\langle N \rangle=11.35$). 

The collapse for both DNs and sacrificial SNs shows that the maximum stress at the first peak is determined by the strength of the sacrificial network, resulting in the linear scaling between $\sigma_{\text{max}}$ and $\lambda_0^{-2}$ shown in Fig.~\ref{fig:fig3}(b). In experiment, a similar scaling was found for the yield stress in ductile DNs~\cite{Millereau2018}, implying that in experiment the yield stress is determined by the strength of the sacrificial network. However, unlike our simulations this linear scaling is only found at high $\lambda_0$. At low $\lambda_0$, the experimentally measured yield (or breaking) stress increases with $\lambda_0$.~\cite{Millereau2018} Our explanation for this experimental observation is that the fracture strength of polymer networks is not only determined by the areal strand density, but also by the presence of defects. Because stress concentrates around defects, their presence can drastically reduce the global stress at which macroscopic cracks are formed and global failure is induced. In DNs the effect of these defects in the sacrificial network is mitigated, because the expansion of the defects into macroscopic cracks is hampered by the matrix chains. An increase in $\lambda_0$ increases the volume fraction of matrix chains and thus increases the screening effect. As a result, the yield (or breaking) stress will increase with $\lambda_0$ as long as global failure is induced by defects in the sacrificial network. Only at high $\lambda_0$, when most defects are screened by the matrix chains, the areal strand density will dominate the fracture response, leading to the expected decrease in the yield stress with increasing $\lambda_0$. In our simulations we do not observe this sensitivity to defects, because our networks are too small to contain defects that can dominate the fracture response.

On passing, we note that the post-peak response of the simulated networks is more ductile than for experimental ones. This is a finite-size effect, also observed in elastic spring networks~\cite{Dussi2020}. In the remainder, we therefore focus on the pre-peak behavior, which we have shown to be consistent with the experimental observations. Furthermore, we note that networks formed at a higher number density of $\rho=0.34$ behave in a similar way to networks formed at $\rho=0.15$ (see Supporting Information for details).

\subsection{Deviations from affine deformation}

The goal of this work is to assess whether the process of inhomogeneous load redistribution affects the global response and the process of damage accumulation at the local level. To quantify this effect, we compare the results of our simulations, where load sharing is intrinsically captured, with an affine prediction for the response, which by definition does not take into account the interaction between the networks. The simulations more closely resemble the experimental reality, where the local load distribution is a result of inter-network rearrangements and excluded volume interactions between networks, while the affine prediction assumes that the local deformation exactly follows the globally applied deformation implying that interactions within and between networks have a negligible effect on the global response. We make these affine predictions based on the evolution of the distribution in end-to-end distances under affine deformation, similar to the statistical damage mechanics models discussed earlier. For easy comparison between chains of different lengths, we introduce the dimensionless chain-stretch $r/L$ with $r$ the end-to-end distance and $L$ the total chain length. We assume that the polymers break at an average chain-stretch of $r/L=1.129$, which corresponds to the stretch where the activation barrier for bond rupture is equal to the thermal energy~\cite{Wang2019} (see Supporting Information for details). Note that in our simulations for sacrificial SNs chains break around $r/L=1.08$. We attribute this lower value to the presence of topological constraints, which can lead to an underestimation of the chain tension based on the end-to-end distance. Since the contribution of each polymer to the affine stress-strain response is independent, crack nucleation or propagation can not be captured in the affine prediction. Our comparison, therefore, focuses on the damage accumulation regime prior $\sigma_{\text{max}}$.

\subsubsection{Damage accumulation}

\begin{figure}
    \centering
    \includegraphics{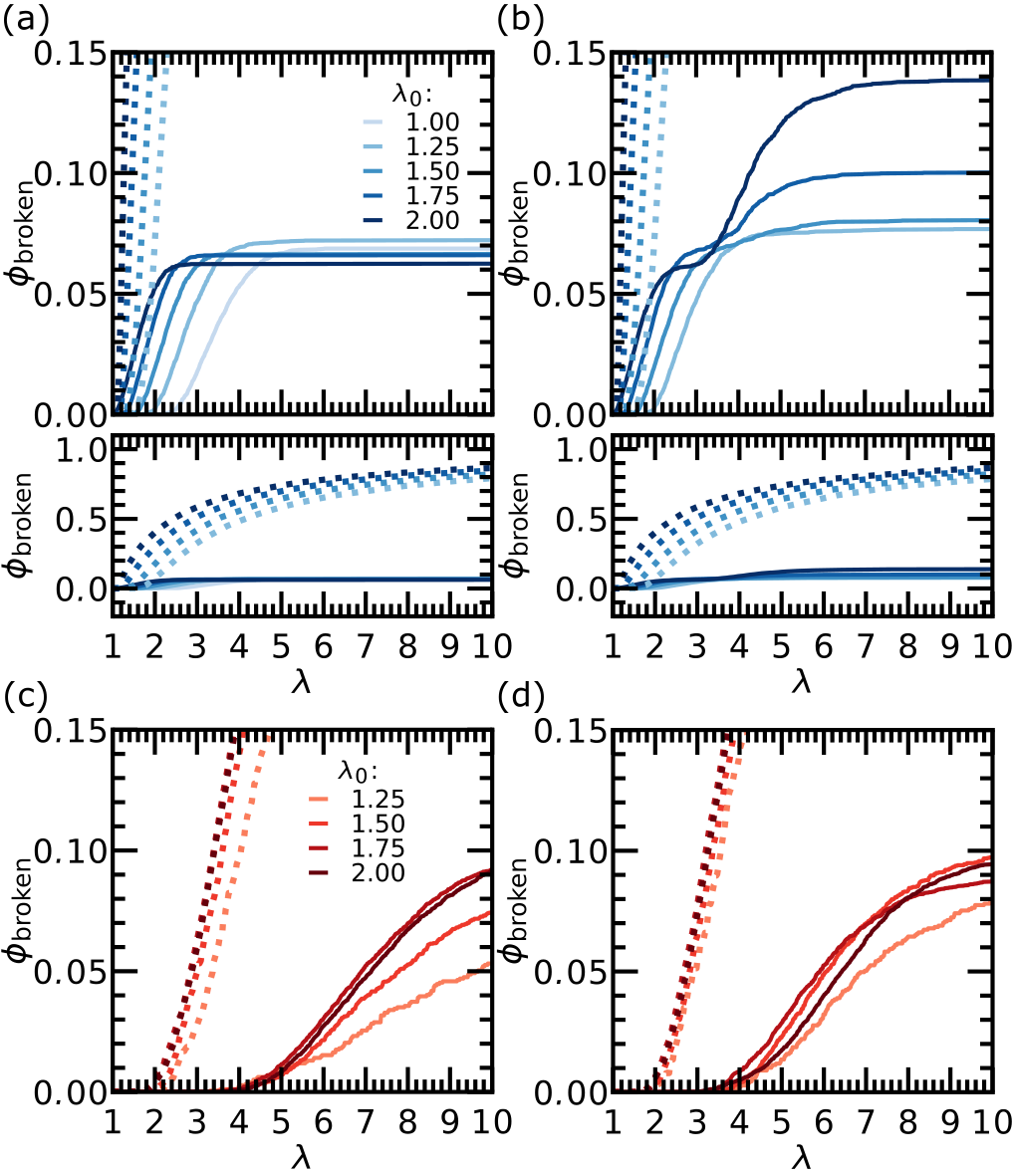}
    \caption{ Fraction of broken chains $\phi_{\text{broken}}$ as a function of global stretch. The affine prediction for $\phi_{\text{broken}}$ is indicated with the dashed lines. (a) $\phi_{\text{broken}}$ for sacrificial chains in the sacrificial SN ($c_1=5\%$) for a range of $\lambda_0$ as indicated in the legend. To illustrate the large difference between the affine prediction and the simulations results, we plot the curves for the entire range of $\phi_{\text{broken}}$ in the lower panel and for a smaller range in the upper panel. (b) $\phi_{\text{broken}}$ for sacrificial chains in DNs ($c_1=5\%$, $c_2=1\%$) for the same swelling ratios as (a). (c) $\phi_{\text{broken}}$ for matrix chains in matrix SNs ($c_2=1\%$) for a range of $\lambda_0$ as indicated in the legend. (d) $\phi_{\text{broken}}$ of matrix chains in DNs ($c_1=5\%$, $c_2=1\%$) for the same swelling ratios as (c).}
    \label{fig:fig4}
\end{figure}

From the microscopic point of view, the main characteristic of the double network response is the enhanced fracture of sacrificial chains. Indeed, we observe a significant increase in the fraction of broken sacrificial chains $\phi_{\text{broken}}$ in the DN (Fig.~\ref{fig:fig4}(b)) compared to the SN (Fig~\ref{fig:fig4}(a)). The enhancement increases with $\lambda_0$, reaching up to a \SI{100}{\%} increase for $\lambda_0=2.00$. This observed enhancement is a clear indication that in our simulations on DNs the sacrificial network interacts with the matrix network.

Comparing the simulation data (solid lines) with the affine predictions (dashed lines), we find that the affine prediction overestimates $\phi_{\text{broken}}$ by almost an order of magnitude for both the sacrificial SN and sacrificial DN (see bottom panels in Fig.~\ref{fig:fig4}(a) and (b)). In the affine prediction, the behavior of all chains in a single network is considered to be independent, i.e., if a chain breaks, this has no effect on the stress carried by the neighboring chains. The significant overestimation of  $\phi_{\text{broken}}$ by the affine models, implies that in our simulation interactions at the network level play an important role in the failure response. The simplest way to introduce network structure into the affine model, would be to consider global failure when percolation is lost. However, in such a simple model the fraction of broken chains is still too high with respect to our simulations ($\phi_{\text{broken}}\approx0.99$).~\cite{Li2020a} In fact, the fraction of broken chains observed in our simulations is closer to the fractions observed for failure of athermal elastic networks,~\cite{Deogekar2019,Dussi2020} where the fracture response is controlled by network rigidity.~\cite{Driscoll2016,Berthier2019,Dussi2020,Tauber2020}

Going back to our simulation data we also find that the rate of chain failure (the slope of the curves) drops significantly at the start of the transition regime in the stress-strain curve for DNs (Fig.~\ref{fig:fig3}(c)), implying that in a DN the fracture of sacrificial chains takes place in two steps.  This is in sharp contrast to the affine prediction where the development of $\phi_{\text{broken}}$ is the same for both the sacrificial SN and the sacrificial DN, due to the absence of interactions between the two networks. 

Combining these insights, we hypothesize that the two-step fracture mechanism in our simulations is controlled by network interactions. The first step is controlled by interaction within the sacrificial network, while the second step is controlled by the topological constraints between the sacrificial network and the matrix. A drop in the fracture rate has been observed experimentally for elastomers~\cite{Millereau2018} and also experiments on hydrogels identified more than one fracture regime~\cite{Nakajima2013a}. Finally, we note that in our simulations the rupture of matrix chains does not take place in two steps; however, the fracture of matrix chains occurs earlier in DN networks than in the matrix SN (Fig.~\ref{fig:fig4}(c) and (d)).

\subsubsection{Which chains are likely to break?}

\begin{figure}
    \centering
    \includegraphics{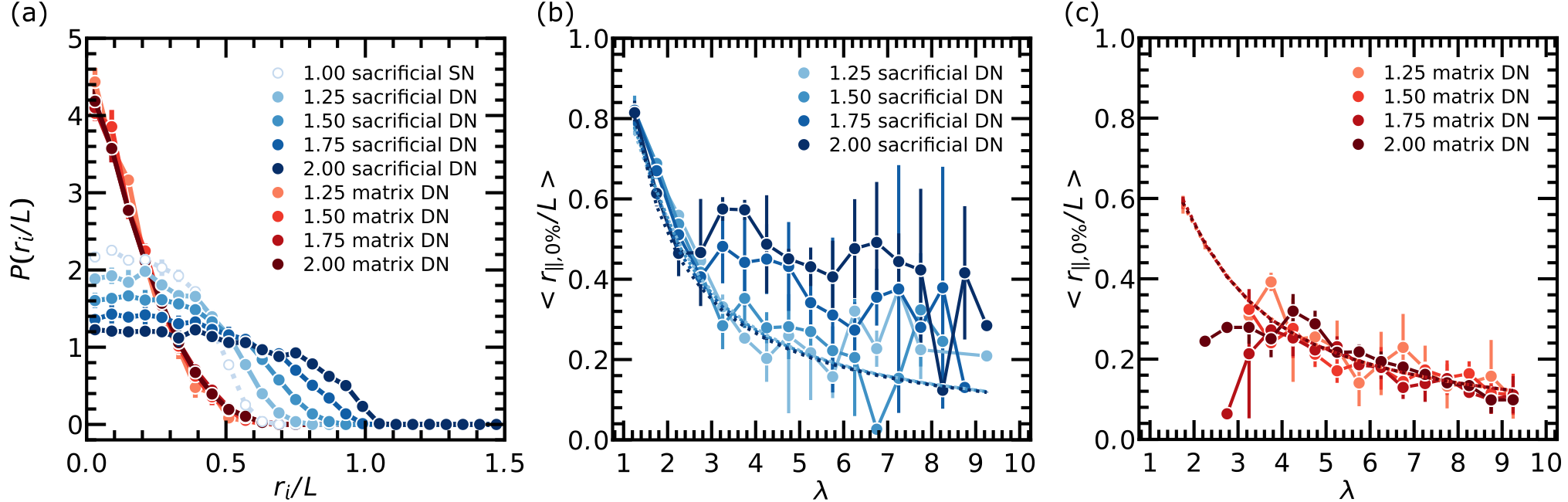}
    \caption{ Role of initial chain-stretch, or pre-stretch, in chain scission. (a) Chain pre-stretch distribution at 0\% strain for swollen sacrificial networks embedded in a matrix (blue) as well as for matrix networks formed inside swollen sacrificial networks (red). The swelling ratios $\lambda_0$ are indicated in the legend. In addition to the data for DNs (solid symbols) the pre-stretch distribution of the sacrificial network prior to swelling ($\lambda_0=1.00$) is shown (open symbols).  Here the stretch $r_i/L$ is the end-to-end distance in one direction divided by the contour length of the polymer. The distributions in these plots are based on $r_i/L$ in all three dimensions. (b) and (c) Average initial chain-stretch along the direction of applied strain $\langle r_{\parallel,\text{0\%}}/L \rangle$ of broken chains as a function of the global strain $\lambda$ at chain rupture for (b) sacrificial DNs and (c) matrix DNs. The dashed lines represent the affine prediction. In all plots the error bars indicate the standard deviation over 4 configurations. }
    \label{fig:fig5}
\end{figure}

A next step in studying the microscopic fracture response is to ask whether we can predict which chains will break. A good first guess would be that shorter chains are likely to break at a lower global strain than longer ones, because for an ideal chain the average stretch at break scales as $Nb/(\sqrt{N}b) = \sqrt{N}$. However, we do not find this trend for the sacrificial network in either the simulations results or the affine prediction (see Supporting Information). This is because in a network the average end-to-end distance of a polymer is constrained by the connections with other chains in the network. This results in a distribution in the average chain-stretch, or pre-stretch, especially in disordered networks with a distribution in chain-length and local connectivity. In Fig.~\ref{fig:fig5}(a) we show the distribution in pre-stretch as the distribution in $r_i/L$, which is the end-to-end distance along one axis, divided by the contour length $L$ of the polymer. 

Considering this distribution in average pre-stretch we could hypothesize that instead of the chain-length the chain-stretch at $\SI{0}{\percent}$ strain determines when a chain will break, so that the sequence in which bonds break can be predicted from the initial chain-stretch distribution. This is also assumed in the statistical damage mechanics approach~\cite{Vernerey2018}. In Fig.~\ref{fig:fig5}(b) and (c) we plot the average pre-stretch at $0\%$ strain $\langle r_{\parallel,\text{0\%}}/L\rangle$ of broken chains as a function of the global stretch $\lambda$ at which the chains break including both the simulation results (solid lines) and the affine prediction (dashed lines).  For the sacrificial network we indeed find that at low strains, the initial chain-stretch does scale with the global strain at break, just as for the affine prediction. For SNs this is true for almost all broken chains (see Supporting Information). However, in DNs this correlation becomes weaker with an increasing $\lambda_0$ and for high $\lambda_0$ the correlation even seems to be lost after the first peak stress (this point also corresponds to the minimum in the chain rupture rate). This observation indicates that up to the peak stress the breaking of chains is largely defined by the configuration at 0\% strain. i.e. network rearrangements do not affect the tension on the chains that break before the peak. However, after this peak stress (the start of the transition region in the DN) the initial structure no longer controls which sacrificial chains break. As a consequence, the initial chain-stretch is not a predictor of failure anymore and the interactions with the matrix (i.e. topological constraints) dominate. This interpretation aligns with our hypothesis that network fracture takes place in two steps.

Note that the initial chain-stretch distribution of the sacrificial networks is determined by the structure of the network and the level of swelling $\lambda_0$. In our simulations the evolution of chain-stretch with $\lambda_0$ is largely affine with respect to the distribution at $\lambda_0=1.00$ (see Supporting Information). We also find that the initial chain-stretch distribution of both the sacrificial and matrix networks are the same in the SN and the DN, indicating that prior to deformation, interactions between the networks are negligible. Note that for the matrix polymers we do find a correlation between chain-length and breaking strain in both the simulations results and the affine prediction; this is because there is a wider distribution in chain-lengths in the matrix networks (see Supporting Information).

\subsubsection{Which chains do break?}

\begin{figure}
    \centering
    \includegraphics{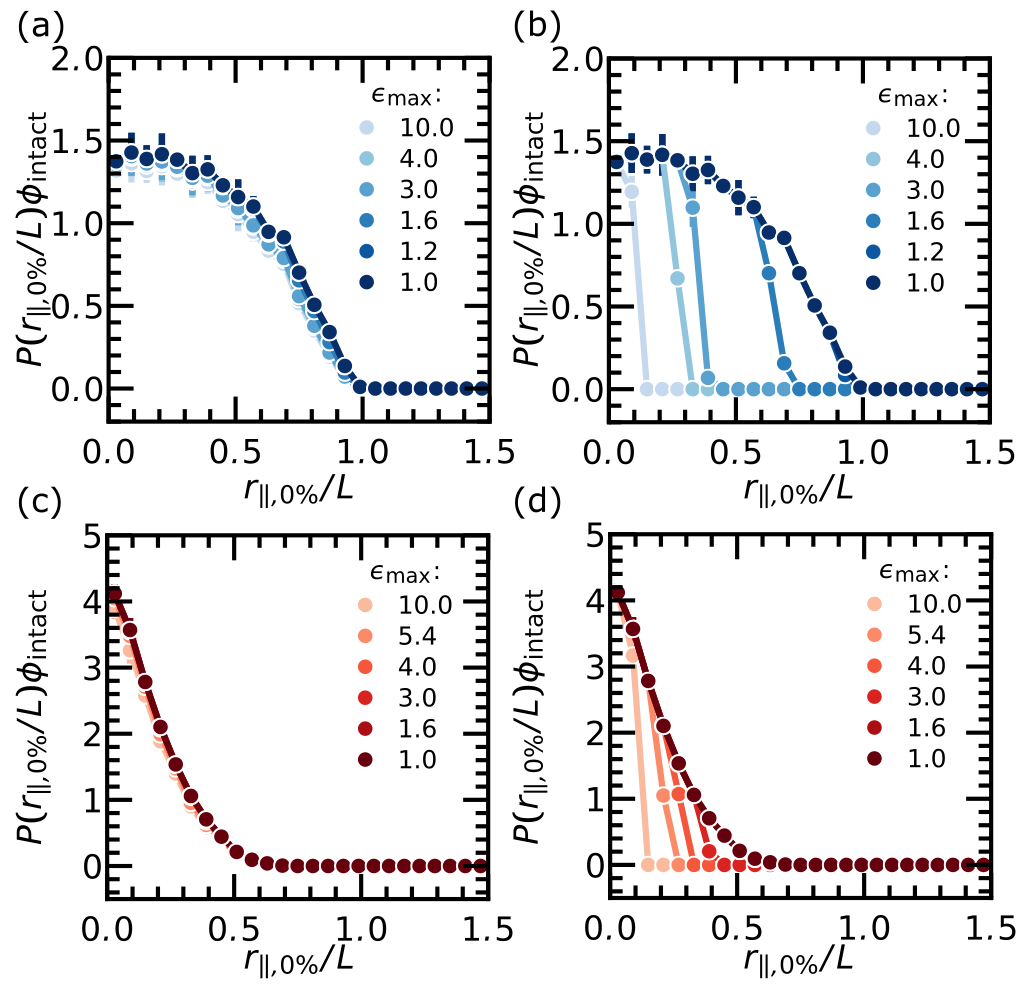}
    \caption{ Predictability of chain rupture based on initial average chain-stretch. (a) Distribution in initial chain-stretch for all intact sacrificial chains in a DN at $\lambda_0=1.75$ at a maximum strain $\epsilon_{\text{max}}$. The distributions are scaled by the fraction of intact chains with respect to the initial configuration. (b) Affine prediction for the sacrificial chains in the DN. (c) Distribution in initial chain-stretch for all intact matrix chains in a DN at a maximum strain $\epsilon_{\text{max}}$. (d) Affine prediction for matrix chains in a DN. }
    \label{fig:fig6}
\end{figure}

The observation that the initial chain-stretch is predictive for when a bond breaks over a large strain-range (Fig.~\ref{fig:fig5}), similar to the affine prediction, is surprising considering the enormous overestimation of broken chains by the affine approximation (Fig.~\ref{fig:fig4}). To investigate what is going on, we plot the distribution of initial chain-stretch for all the intact chains at a particular strain (Fig.~\ref{fig:fig6}). We find that although the initial chain-stretch is predictive for when a chain can break, this does not mean that all chains with that initial chain-stretch will break. Actually, only few of those chains break, which is in sharp contrast with the affine prediction Fig.~\ref{fig:fig6}(b). Our explanation is that in the first failure regime the deformation is largely affine in the undamaged network, but once a polymer breaks, significant stress-relaxation becomes possible via rearrangements at the local level, alleviating the tension on polymers that surround the broken chain. In other words, we expect that stress heterogeneity within the network grows once damage starts to accumulate. Similar behavior is observed for the rupture of matrix chains as shown in Fig.~\ref{fig:fig6}(c) and Fig.~\ref{fig:fig6}(d).

\subsubsection{Evolution of the pre-stretch distribution}

\begin{figure}
    \centering
    \includegraphics{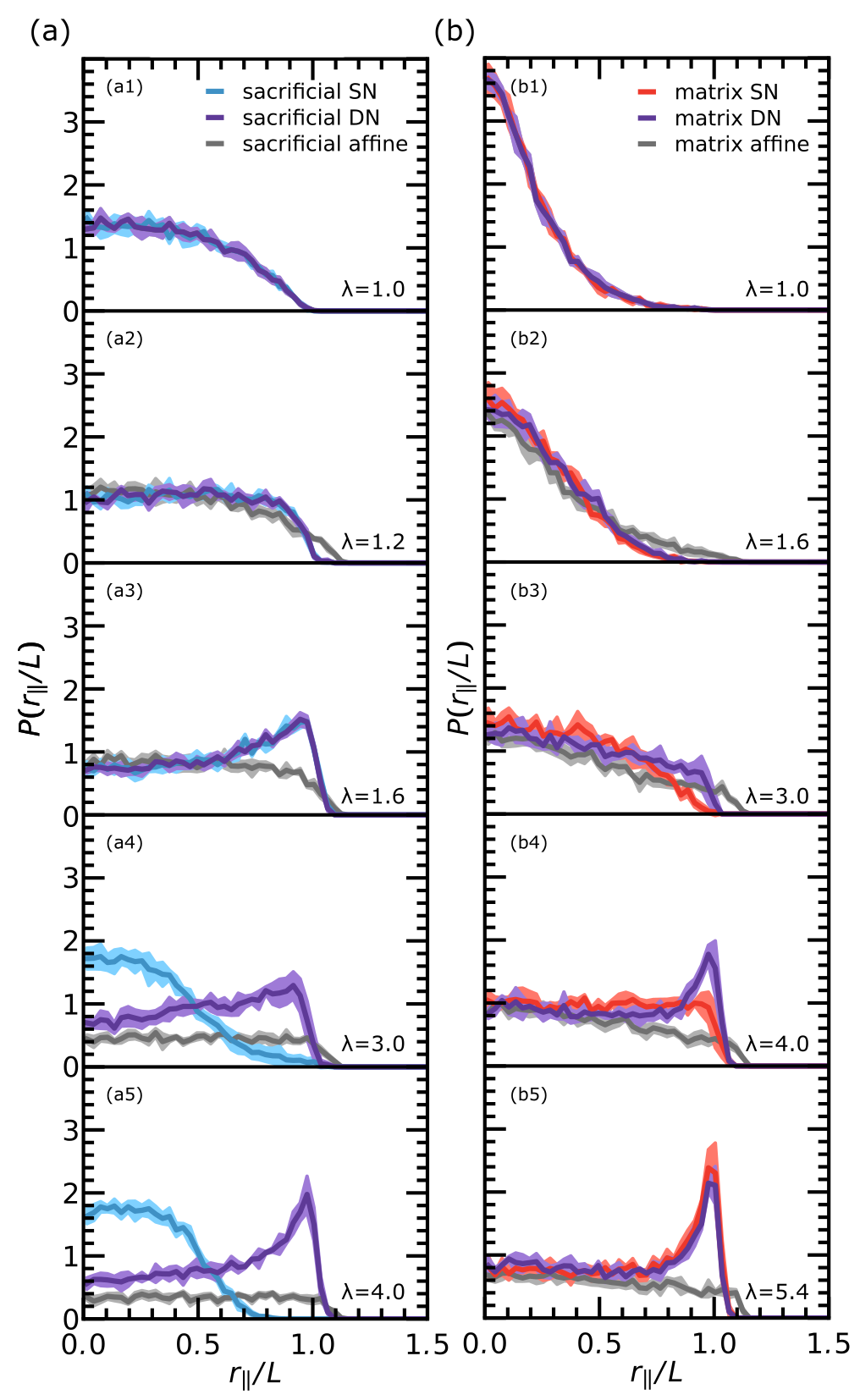}
    \caption{ Evidence of network-network interaction based on stretch distribution. Comparison of chain-stretch in a network (SN and DN) at different strains as indicated in the plot ($\lambda_0=1.75$). (a) The sacrificial network as stand-alone network (SN, blue) and as part of a DN (purple). (b) Matrix network as a stand-alone network (SN, red) and as part of a DN (purple). For both panels the affine prediction for the chain-stretch distribution is indicated in gray. The shading indicates the standard error computed over the 4 configurations. }
    \label{fig:fig7}
\end{figure}

The data on the accumulation of damage suggest that in our simulations inhomogeneous redistribution of stress is taking place both within single networks during the first failure process and between networks during the second failure process. As the stress distribution within a polymer network is strongly related to the distribution in chain-stretch, we expect that any inhomogeneous mode of stress-redistribution should be reflected in the evolution of the chain-stretch distribution as a function of strain. Furthermore, by comparing the evolution of chain-stretch in our simulations, where intra and inter-network interactions are accounted for, with the affine prediction for the evolution of chain-stretch, where intra and inter-network interaction are neglected, we can identify if and when network level processes affect the mechanical response of polymer networks.

In particular, we look at the distribution in end-to-end distances parallel to the axis of deformation $r_{\parallel}/L$ (Fig.~\ref{fig:fig7}), as the chain-stretch along the direction of applied deformation is primarily responsible for the global mechanical response. 

When the strain is increased from $\lambda=1.0$ to $\lambda=1.2$ for a network at $\lambda_0=1.75$ we see that the distribution for both the sacrificial SN and the sacrificial DN flattens, in a similar way as expected for affine deformation (Fig.~\ref{fig:fig7}(a2)). Around the first peak stress at $\lambda=1.6$ (Fig.~\ref{fig:fig7}(a3)) we see an accumulation of chains that are stretched up to their contour length ($r_{\parallel}/L=1.0$) both in the sacrificial SN and sacrificial DN. This behavior is very different from the response expected based on affine deformation, where no peak is visible, and implies that in our simulations additional load imposed on strongly stretched chains does not always lead to chain rupture (as in the affine prediction), but can also lead to redistribution of load to less stretched polymer chains. This mode of inhomogeneous stress redistribution might be caused by local stress relaxation after chain rupture, as discussed in the previous section. As the same trend is observed in sacrificial SNs and sacrificial DN chains this is a clear sign of inhomogeneous redistribution of load within the network. 

At higher strains ($\lambda=3.0$ and $\lambda=4.0$, Fig.~\ref{fig:fig7}(a4) and (a5)) the behavior of sacrificial chains in the DN diverges from the behavior in SNs. While the sacrificial chains in the SN relax to a stretch below the initial chain-stretch and remain there at higher strains as a result of macroscopic network fracture, a large part of the sacrificial chains in the DN remain close to their entropic stretching limit ($r_{\parallel}/L=1.0$). At $\lambda=4.0$ (Fig.~\ref{fig:fig7}(a5)) the number of chains at maximum extension increases again in the DN. The divergence between the SN and DN chain-stretch demonstrates the effect of adding a matrix on the microscopic stress distribution within the network, revealing that due to inter-network interactions, sacrificial chains are still under significant tension beyond the first peak stress. Only far beyond the second peak in stress ($\lambda=10.0$) the sacrificial chains start to relax, as strongly stretched chains rupture (not shown), due to macroscopic fracture of the DN.

In the matrix network (Fig.~\ref{fig:fig7}(b)) we observe that initially the SN and DN behave in the same way, but between a stretch of $\lambda=\num{3}$ and $\lambda=\num{4}$ we see that there are more stretched chains in the DN compared to the SN, indicating that during the transition regime an interaction between the two networks arises. These data also match the shift in the stress response we have seen earlier for the matrix DN compared to the matrix SN (Fig.~\ref{fig:fig2}). Overall, it becomes clear that in our simulations we find inhomogeneous stress redistribution at the network level at low strains and inhomogeneous stress redistribution between networks at high strains. These different processes for stress management explain the two distinct failure regimes identified for the sacrificial chains in Fig.~\ref{fig:fig4}.

\subsubsection{Impact on global response}

\begin{figure}
    \centering
    \includegraphics{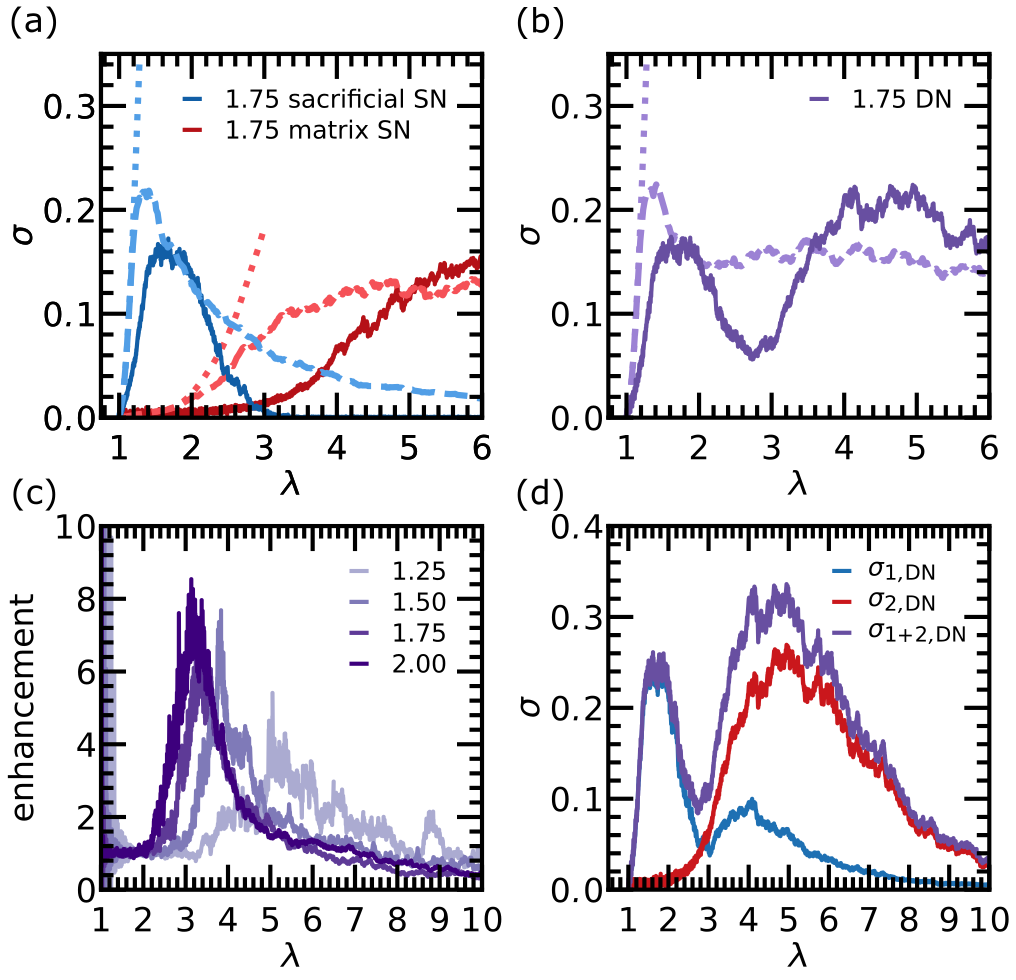}
    \caption{The effect of load sharing on the global stress response. (a) and (b) Comparison of the stress response from simulation with the affine prediction ($\lambda_0=1.75$, $c_1=5\%$, $c_2=1\%$). The stress response upon affine deformation is determined from the average crosslinker positions at 0\% strain, assuming that the stress carried by the polymer in between the crosslinkers can be described as a freely-jointed chain with extensible quartic bonds (eFJC, see method section for details). (a) SN response of the sacrificial network (blue) and the matrix network (red) from simulation (solid line), the affine prediction for the mechanical response (dotted line) and the affine prediction considering polymers with $r/L>1.129$ to be broken (dashed line). (b) DN response. (c) Enhancement in measured stress in the DN $\sigma_{1+2,\text{DN}}/(\sigma_{1,\text{SN}} + \sigma_{2,\text{SN}})$ with $\sigma_{1+2,\text{DN}}$ the measured stress in the DN, $\sigma_{1,\text{SN}}$ the measured stress in the sacrificial SN and $\sigma_{2,\text{SN}}$ the measured stress in the matrix SN. Swelling ratios are indicated in the legend. (d) Stress response of the sacrificial network $\sigma_{1,\text{DN}}$ (blue) and the matrix network $\sigma_{2,\text{DN}}$ (red) embedded in the DN at $\lambda_0=1.75$. For reference we also plot the total stress response of the DN $\sigma_{1+2,\text{DN}}$ (purple). }
    \label{fig:fig8}
\end{figure}

So far, we have shown that the microscopic response of the networks is dominated by processes of inhomogeneous stress redistribution, both within single networks and between networks. With increasing strain, these microscopic processes deviate further from the affine picture. The question that remains is: do these processes only matter at the local level or do they also affect the global stress response? To answer this question, we make a prediction for the stress response under affine deformation, based on the pre-stretch distributions shown previously, assuming that the resistance to deformation of a single chain can be described as an extensible freely jointed  chain (eFJC, see method section for details). Although our short chains are not expected to behave exactly as ideal chains, a cross-check of this method with the simulation result reveals that this assumption still serves our purpose (see Supporting Information for details).

Comparing the affine prediction with the simulation results, we find that in the linear regime, the affine prediction agrees quite well with the simulations at low swelling ratios (see Supporting Information). For the response at larger strains we look at the networks for $\lambda_0=1.75$ in Fig.~\ref{fig:fig8}(a) and Fig.~\ref{fig:fig8}(b). We immediately see that in the affine prediction (dotted lines) strain-stiffening sets in earlier than in our simulations (solid lines) both for the SNs (Fig.~\ref{fig:fig8}(a)) and the DN (Fig.~\ref{fig:fig8}(b)). This suggests that in the non-linear elastic regime network rearrangements reduce the tension on individual polymers. Even though the impact on the global stress is significant, calculation of the non-affine displacement of the crosslinks shows that in most cases these rearrangements are relatively small in the non-linear elastic regime (see Supporting Information). Because of the non-linear stress response of single polymer chains a small rearrangement can still have a significant impact on stress, providing a possible explanation for the strong effect observed here.

We find that the stress of the affine prediction quickly overshoots the simulation response if bond breaking is not considered for both SNs and DNs (dotted lines). If instead, we assume that chains with a stretch larger than \num{1.129} break, we find an affine SN response (dashed lines) in Fig.~\ref{fig:fig8}(a) that is qualitatively similar to the simulation results (solid lines). However, in the affine case the onset of strain-softening occurs earlier, the maximum stress is higher, the strain at maximum stress is lower and the stress drop after the maximum stress is smoother. These differences imply that inhomogeneous redistribution of load within a network upon breaking of chains has a significant impact on the global stress response. The picture is the same if we compare the affine prediction for the behavior of the DN (dashed line) with the simulation result (solid line) in Fig.~\ref{fig:fig8}(b). We note that the loop observed in the affine case is less pronounced than in the simulations. We attribute this to the broad distribution in chain-length which causes a relatively smooth decay of the affine stress in the sacrificial network (Fig.~\ref{fig:fig8}(a)) and the relatively small difference in crosslinker density between the sacrificial network and the matrix.

We have found several indications from the microscopic response in our simulations that interactions arise between the two networks if the strain is high enough (Fig.~\ref{fig:fig4}, Fig.~\ref{fig:fig5},Fig.~\ref{fig:fig7}). To check if these network-network interactions affect the global stress response we compare the stress of the DN with the sum of the stresses for the individual SNs via the enhancement factor $\sigma_{1+2,\text{DN}} / (\sigma_{1,\text{SN}} + \sigma_{2,\text{SN}})$ (Fig.~\ref{fig:fig8}(c)). We observe that significant enhancement starts after a certain stretch and that the onset of enhancement decreases as a function of $\lambda_0$. The onset of enhancement seems to coincide with the peak strain of the sacrificial SN (Fig.~\ref{fig:fig2}). After the onset of enhancement, the enhancement factor increases up to a factor of \num{8.0} at the peak. Also the location of this peak decreases as a function of $\lambda_0$. The enhancement peaks just after the end of the transition region. If we plot the stress of the sacrificial network and the matrix network in the DN ($\sigma_{1,\text{DN}}$ and $\sigma_{2,\text{DN}}$ in Fig.~\ref{fig:fig8}(d)) together with the total stress $\sigma_{1+2,\text{DN}}$, we see that the enhancement in stress contains contributions of both the sacrificial network and the matrix, further confirming that the enhancement is caused by the interaction between the two networks. The second peak in stress in the sacrificial DN response suggests that even after the yield stress, sacrificial chains resist deformation. Indeed, we find that around this second peak in stress a considerable fraction of the sacrificial chains is stretched up to their contour length (Fig.~\ref{fig:fig7}(a5)).

\section{Conclusions and outlook}
Our simulations confirm that in a DN, both the local and global response is governed by sharing of load at the network level. Similar to experiments~\cite{Millereau2018}, the mechanical response and accumulation of damage at low strains is governed by the sacrificial network even in the strain-softening regime. The behavior at the microscopic level reveals that upon deformation and damage accumulation stress is redistributed inhomogenously within the network via small and local non-affine rearrangements. 

After the yield stress, the mechanical response is controlled by both the sacrificial network and the matrix network. The enhancement in broken chains (Fig.~\ref{fig:fig4}), the change in failure mechanism (Fig.~\ref{fig:fig5}) and the altered stretch distributions (Fig.~\ref{fig:fig7}) reveal that both networks interact with each other through their topological constraints, leading to large non-affine rearrangements at the network level. These inter-network interactions cause an enhancement in the fraction of broken sacrificial chains, in line with the sacrificial bond principle. At the global level these interactions cause a significant enhancement in the stress response of DNs compared to the SNs (Fig.~\ref{fig:fig8}).

The comparison of our simulation data with affine predictions suggests that for any polymer network (SN or DN) the inhomogeneous redistribution of load through the network can be an important mechanism in the non-linear elastic and fracture response. In the non-linear elastic regime, non-affine rearrangements appear to be small. Therefore, we expect that in this regime the rearrangements are mainly driven by the non-linear stress response of entropic springs in combination with the disordered structure of the network. However, in the fracture regime we find strong deviations from the affine prediction. In particular, we observe that a significant fraction of the chains is stretched beyond their contour length (Fig.~\ref{fig:fig7}).  Experiments on single polymer chains also reveal that extension of polymer chains up to this limit is possible.~\cite{Wang2019} These data suggest that in the fracture regime enthalpic stretching could play an important role in the behavior of networks prior to the propagation of a macroscopic crack. 

Our simulations, provide predictions for the effect of load sharing on the microscopic fracture response. Several of these predictions can be verified in experiment. For example, the low fraction of broken chains (Fig.~\ref{fig:fig4}) could be investigated by quantification of the fraction of broken chains prior to crack propagation by incorporating chain scission reporters in the network such as dioxetane~\cite{Ducrot2014} or anthracene~\cite{Slootman2020}. Our simulations also suggest that the distribution in chain-stretch provides information on the (inhomogeneous) redistribution of load within a network and between networks (Fig.~\ref{fig:fig5} and Fig.~\ref{fig:fig7}). Although tracking the evolution of the chain-stretch distribution would be a challenging endeavor, experiments on FRET-based force-sensors shows that this might be possible experimentally~\cite{VandeLaar2018}.

\section{Methods section} 
\subsection{\textit{In silico} synthesis of double network}

The networks are formed by the self-assembly of binary mixtures of bifunctional and tetrafunctional patchy particles as done in Refs.~\cite{Gnan2017,Rovigatti2018,Sorichetti2021}. In order to build the first network we simulate the binary mixture at a number density $\rho_{\text{init}} = 0.17$. We stop the simulation when most ($> 99.9\%$) of the bonds have formed, after which we remove the few clusters that are not attached to the largest one. No more than $3\%$ of the particles are removed at this stage. We take the resulting system, locate all the chains, defined as clusters of bifunctional particles connecting the crosslinkers, and add five monomers to each in order to make the system more swellable (resulting in a number density $\rho_{\text{init,add}} = 0.33$).

In order to swell the network in LAMMPS~\cite{Plimpton1995}, we convert the network of patchy particles to a network of harmonic bonds and equilibrate the network in the NVT ensemble for $10\tau$. Subsequently, we convert the harmonic bonds to quartic bonds and equilibrate the network in the NPT ensemble for $100\tau$ such that the network settles at an equilibrium box size $L_{\text{box},0}$. The resulting network is the sacrificial SN at $\lambda_0=1.00$ and number density $\rho=0.15$. This network is swollen isotropically (NVT ensemble) in steps of $\sim \SI{0.1}{\percent}$ strain  such that the new box size is $L_{\text{box}}=L_{\text{box},0}\lambda_{0}$, providing the sacrificial SNs at higher swelling ratios.

To form the corresponding DNs, we add matrix monomers, and subsequently perform the same self-assembly procedure described above, with the difference that this time the bifunctional and tetrafunctional particles are embedded in the existing polymer network. After the assembly of the binary mixture completes we once again remove the few disconnected clusters but this time we do not add any additional monomers to the chains. The resulting DNs are NVT equilibrated in LAMMPS first using harmonic bonds (for $10\tau$), then using quartic bonds (for $10\tau$). Finally, the matrix SNs are obtained by removing the sacrificial network from the DNs.

For all LAMMPS simulations the time step for integration $dt = 0.001\tau$. For simulations performed in the NVT ensemble the temperature is controlled via a Nos\'{e}-Hoover thermostat and kept fixed at $T=1.0$ (in reduced units) with a damping time $t_{\text{damp}}=0.1\tau$ (\num{100} time steps). In addition, for simulations performed in the NPT ensemble the pressure is fixed at $P=1.0$ (in reduced units) and the corresponding $t_{\text{damp}}=1.0\tau$ (\num{1000} time steps).  

The interaction between the particles is described by the Weeks-Chandler-Andersen (WCA) potential, a truncated version of the Lennard-Jones potential,
\begin{equation}
    U(R) = 4\epsilon \left[ \left(\frac{\sigma}{R}\right)^{12} - \left(\frac{\sigma}{R}\right)^{6}\right] \;\;\;\;\;\;\;\; R<R_c \;\;,
\end{equation}
\noindent with $R$ the inter-particle distance, $\sigma=1.0$ the particle diameter, $\epsilon=1.0$ the depth of the potential well and $R_c=2^{1/6}$ the cut-off distance, unless the particles are connected by a bond, in that case the particle-particle interaction is described by a quartic potential ($K=2351$, $B_1=-0.7425$, $B_2=0.0$, $R_c=1.5$, $U_0=92.74467$).
\begin{equation}
    U(R) = K(R-R_c)^2(R-R_c-B_1)(R-R_c-B2) + U_0 + 4\epsilon \left[ \left(\frac{\sigma}{R}\right)^{12} - \left(\frac{\sigma}{R}\right)^{6}\right] + \epsilon
\end{equation}
\noindent These parameters have been used previously to study polymer rupture.~\cite{Ge2013} The quartic bonds break irreversibly if their extension exceeds $R_c=1.5$. However, the maximum force is already reached around an extension of $1.133 \sigma$ and the bonds are expected to break even earlier, around an extension of $1.08 \sigma$, due to thermal fluctuations (see Supporting Information for details). In some equilibration steps harmonic bonds are used instead of a quartic bond ($K=1000$, $R_0=0.96$).
\begin{equation}
    U(R) = K(R-R_0)^2
\end{equation}

\subsection{Extension protocol}
The stress response is obtained by performing a continuous uniaxial extension at a strain-rate $\dot{\epsilon} = \num{1e-4}$, while keeping the volume of the simulation box constant (lateral dimensions are reduced during extension). A similar procedure has been followed in literature \cite{Wang2017,Yin2020}. Decreasing the strain-rate by a factor 10 does not significantly alter the mechanical response. The stress response $\sigma$ is determined from the virial stress excluding kinetic contributions, which are nevertheless negligible.  First, we calculate the deviatoric (true) stress as $\sigma_{T}= \sigma_{ii} - \sigma_{\text{hydr}}$. Subsequently, we convert this to the engineering stress $\sigma = \sigma_{T}/\lambda$. For every configuration the deformation protocol is performed in the $x$, $y$, and $z$ direction and the output is averaged. Data presented in the manuscript are averages over \num{4} configurations. If error bars are used, they indicate the standard deviations in the values between these \num{4} configurations.

\subsection{Analysis of chains}
Polymer chains are defined as the set of particles in between crosslinkers, the latter having connectivity different from \num{2}. Some of the $N_\text{chains}$ in a network are trivial dangling ends, i.e. one of the ends of the chain has functionality of \num{1}, and are indicated as $N_{\text{dang}}$. We also identified (first-order) loops when both chain ends share the same crosslinker, and we indicated these as $N_\text{loops}$. Both loops and dangling ends are expected to not contribute to the mechanical response; therefore, in a first approximation we can define the active chains as $N_{\text{act}}=N_{\text{chains}}-N_{\text{dang}}-N_{\text{loops}}$.  We define the chain-length, $L$, as $b*(N_{\text{beads}}-1)$ where $b=0.96$ is the rest length of the quartic bond and $N_{\text{beads}}$ the number of particles in a polymer chain including the crosslinkers. We define the end-to-end distance $r$ as the Euclidean distance between crosslinkers. To calculate $r$ prior to deformation we use the average crosslinker positions from a simulation run of $\num{10000}\tau$, where the crosslinker locations are saved every $\num{50}\tau$. For the calculation of $r$ during deformation the crosslinker positions are based on snapshots which are saved every $\Delta \lambda=0.01$. In both cases coordinates are unwrapped to correct for periodic boundary crossings and corrected with respect to their combined center of mass. We consider a polymer chain to be broken if one of the bonds inside the chain breaks. Breaking of bonds is reported via a custom extension of the LAMMPS code.

\subsection{Affine predictions}
Affine predictions for $r_i/L$ distributions and stress are made based on the time-averaged positions of chain-ends of active chains, i.e., crosslinkers. Based on this configuration we can determine the average location of chain-ends after affine deformation. From these positions we calculate the end-to-end distances of all the polymer chains. Chains are considered broken if their chain-stretch exceeds the maximum stretch of a quartic bond ($1.08/b = \num{1.129}$). Broken chains are not included in the distributions.

To predict the stress response we assume that the stress response of the single polymers can be described as an extensible freely jointed chain (eFJC)~\cite{Mao2017} which covers both entropic and enthalpic contributions. In this way we can obtain the stress contribution of every polymer based on the location of the chain ends. Combining the contributions of all active polymers, we get our prediction of the virial stress tensor and thus the global response based on affine deformation.

To find the stress contribution of every polymer we rewrite the chain-stretch as $r/L =  \frac{r_{L}}{L} \lambda_b$, where $r_{L}/L$ is the entropic chain-stretch and $\lambda_{\text{b}} = R_{\text{b}}/b$ the enthalpic stretch of a quartic bond. $\lambda_{\text{b}}$ is found by numerically solving the force balance $\frac{dU_{\text{quartic}}(\lambda_{\text{b}})}{d\lambda_{\text{b}}}\lambda_{\text{b}} = k_{\mathrm{B}}T\frac{r}{\lambda_{\text{b}}L }\mathcal{L}^{-1}\left( \frac{r}{\lambda_{\text{b}}L}\right)$ according to Ref.~\cite{Mao2017}, where $\mathcal{L}^{-1}$ is the inverse Langevin equation (we use the approximation by Puso~\cite{Jedynak2015}). Based on this value we can calculate the force from $F = \frac{k_{\mathrm{B}}T}{b\lambda_{\text{b}}} \mathcal{L}^{-1}\left( \frac{r}{\lambda_{\text{b}} L} \right)$. Note that $k_{\mathrm{B}}T=1.0$ in reduced units.

\begin{acknowledgement}
This work is part of the SOFTBREAK project funded by the European Research Council (ERC Consolidator Grant 682782).
\end{acknowledgement}

\begin{suppinfo}

Explanation of parameterization, additional network characterization, mechanical response for networks formed at $\rho=0.34$ and validation of the eFJC estimation for the stress carried by a coarse grained polymer.

\end{suppinfo}

\bibliography{main}

\end{document}


\clearpage

\subsection{Parameterization of the coarse-grained networks}

The input parameters for the network generation process are the particle number density $\rho$, the number of particles $N_1$, and the fraction of crosslinkers $c_1$. $\rho$ and $N_1$ set the initial box size via $L_{\text{box}}=\sqrt[3]{N_1/\rho}$. Our choice for $c_1$ is inspired by elastomeric double networks synthesized from polyethyl acrylate (PEA).\cite{Millereau2018} For these networks the fraction of crosslinkers $c_{\text{cross}}$ is \SI{1.45}{\mol\percent} relative to the monomer. Because we simulate coarse-grained networks, the particles in our simulation do not represent a single monomer, but a Kuhn length $b$ (which typically encompasses several monomers). To convert $c_{\text{cross}}$ to a $c_{1}$, we used the characteristic ratio $C_{\infty}$ as a coarse-graining factor. Assuming the polymer behaves as a freely jointed chain, $C_{\infty}$ is equal to the number of monomers within a Kuhn length. For PEA $C_{\infty}=\num{9.67}$, thus our particle represents \num{9.67} ethyl acrylate monomers.\cite{Millereau2018} Using this information we can determine the crosslinker fraction as $c_1 = c_{\text{cross}}C_{\infty}$, which for PEA gives a crosslinker fraction of approximately 10\%. Because such a high $c_1$ would result in very short chains, we decided to set $c_1=\SI{5}{\percent}$.

\begin{figure}
    \centering
    \includegraphics[scale=0.75]{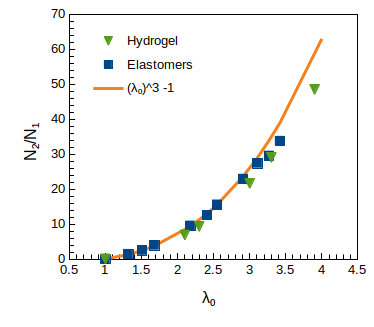}
    \caption{Fraction of matrix network monomers with respect to sacrificial network monomers as a function of $\lambda_0$. The plot show the fractions for experiments on elastomers\cite{Millereau2018} and on hydrogels\cite{Matsuda2016}. The orange line is our approximation $N_2/N_1 \approx (\lambda_0)^3-1$. }
    \label{fig:ch6figS1}
\end{figure}

The parameters for the second network are again inspired by experimental data.\cite{Millereau2018,Matsuda2016} In our model the variables are $\lambda_0$, $N_2$ and $c_2$. Because in elastomeric double networks, the sacrificial network is swollen with matrix monomers, $N_2$ is coupled to $N_1$ and $\lambda_0$. Considering that in an elastomer the monomer density remains constant we can write expressions for the density of the initial network $\rho=\rho_{1,\text{init}}=\frac{N_1}{V}$ and the swollen network $\rho=\rho_{1+2,\text{new}}=\frac{N_1}{V\lambda_0^3} + \frac{N_2}{V\lambda_0^3}$, with $V$ the volume of the system. By equating these expressions, we find that the ratio between $N_2$ and $N_1$ can be written as $N_2/N_1 = \lambda_0^3 -1$. Plotting this relation together with experimental data on $\lambda_0$ in Figure~\ref{fig:ch6figS1}, we find that this relation approximates the experimental situation for double network elastomers. Although swelling in double network hydrogels can be controlled independently from the monomer concentration, we do observe in at least one article on hydrogels~\cite{Matsuda2016} that $N_2/N_1 \approx \lambda_0^3 -1$.

The fraction of crosslinkers in the second network relative to the number of matrix monomers $c_{\text{cross},2}$ is typically a lot lower than in the sacrificial network. For example, in elastomers $c_{\text{cross},2} = \SI{0.01}{\mol\percent}$, which would lead to a $c_2=c_{\text{cross},2}C_{\infty}\approx \SI{0.1}{\percent}$.  For our system size it would be improbable to get a percolating second network under those conditions. Therefore we opt for a $c_{2}$ of $\SI{1}{\percent}$.

\subsection{Chain length distribution}

\begin{figure}
    \centering
    \includegraphics{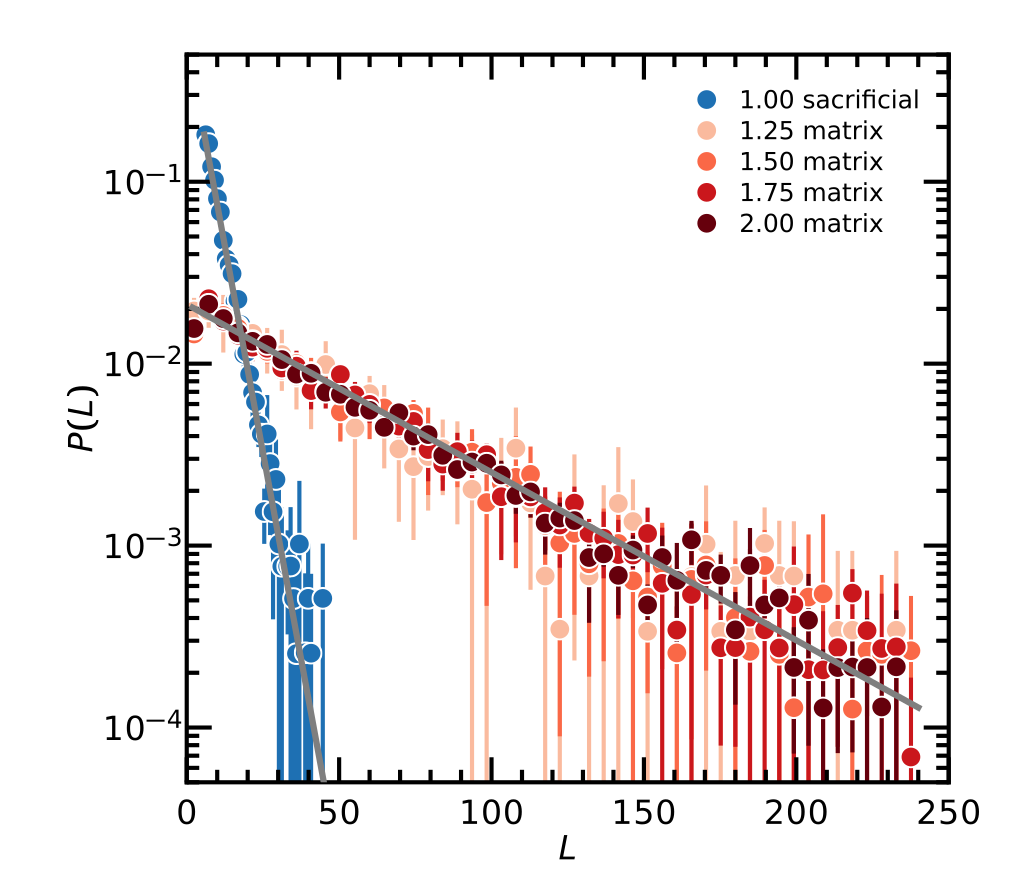}
    \caption{Chain length distribution within the sacrificial network (blue) and within the matrix networks (red) ($\rho=0.15$, $c_1=5\%$, $c_2=1\%$). Error bars indicate the standard deviation over \num{4} configurations. The gray lines indicate the distribution in chain length as expected from asymptotic Flory statistics.~\cite{Rovigatti2018}  }
    \label{fig:ch6figS2}
\end{figure}

Just like in hydrogels~\cite{Nakajima2012} and elastomers~\cite{Ducrot2014} the polymer chains in our \textit{in silico} networks have a distribution in chain-length and pre-stretch. Our \textit{in silico} networks show an exponential distribution in chain length (Figure~\ref{fig:ch6figS2}) for both the sacrificial network and the matrix. This is in accordance with earlier works~\cite{Gnan2017,Rovigatti2018} where it is demonstrated that the chain-length distribution follows asymptotic Flory statistics~\cite{Rovigatti2018} as indicated by the grey lines.

Specifically, we create a network from $n_{\text{mon}} = n_{A+B}$ monomers, of which $n_A$ monomers and $n_B=n_{\text{mon}}c_{\text{x-link}}$ crosslinker. The crosslinkers have a functionality $f_{B}$. If all crosslinkers are used efficiently the number of chain ends $n_e = f_B n_B$ and thus we also know that the number of chains $n_c = f_B n_B/2$ as all chains have two chain ends. The minimum number of particles in a chain $n_{\text{min}}$ is set by two factors. (i) crosslinkers (B) can only form bonds with monomers (A) such that the the smallest chain will always include two crosslinkers and one monomer. (ii) we can add additional particles $n_{\text{add}}$ to every chain after network formation to increase the minimum chain length.
If $n_{\text{add}}>0$ we first form a network with $n_{A+B,\text{init}} = n_{A+B} - n_{\text{add}}n_c$ and $n_A = n_{A+B,\text{init}} -n_{B}$. Please note that the segment length of the polymer $N=l-1$ with $l$ the number of particles in a chain. The number of chains of a certain segment length $n^{\text{asympt}}_N$ in this system can be described according to an asymptotic Flory description.~\cite{Rovigatti2018}

\begin{equation}
    n^{\text{asympt}}_N = n_{A} \left( \frac{p_B}{p_A} \right)^2 \left( \frac{p_A - p_B}{p_A} \right)^{N-n_{\text{add}}-2}
\end{equation}

\noindent Here $p_A = 2 n_A / (2 n_A + f_B n_B)$ and $p_B = 1 - p_A$. To get the probability distribution function for chain length we can write 

\begin{equation}
    P_{\text{asympt}}\left( N ; n_{\text{mon}}, c_{\text{x-link}},f_{B}, n_{\text{add}} \right) = \frac{n^{\text{asympt}}_N}{n_c}
\end{equation}




\begin{figure}
    \centering
    \includegraphics{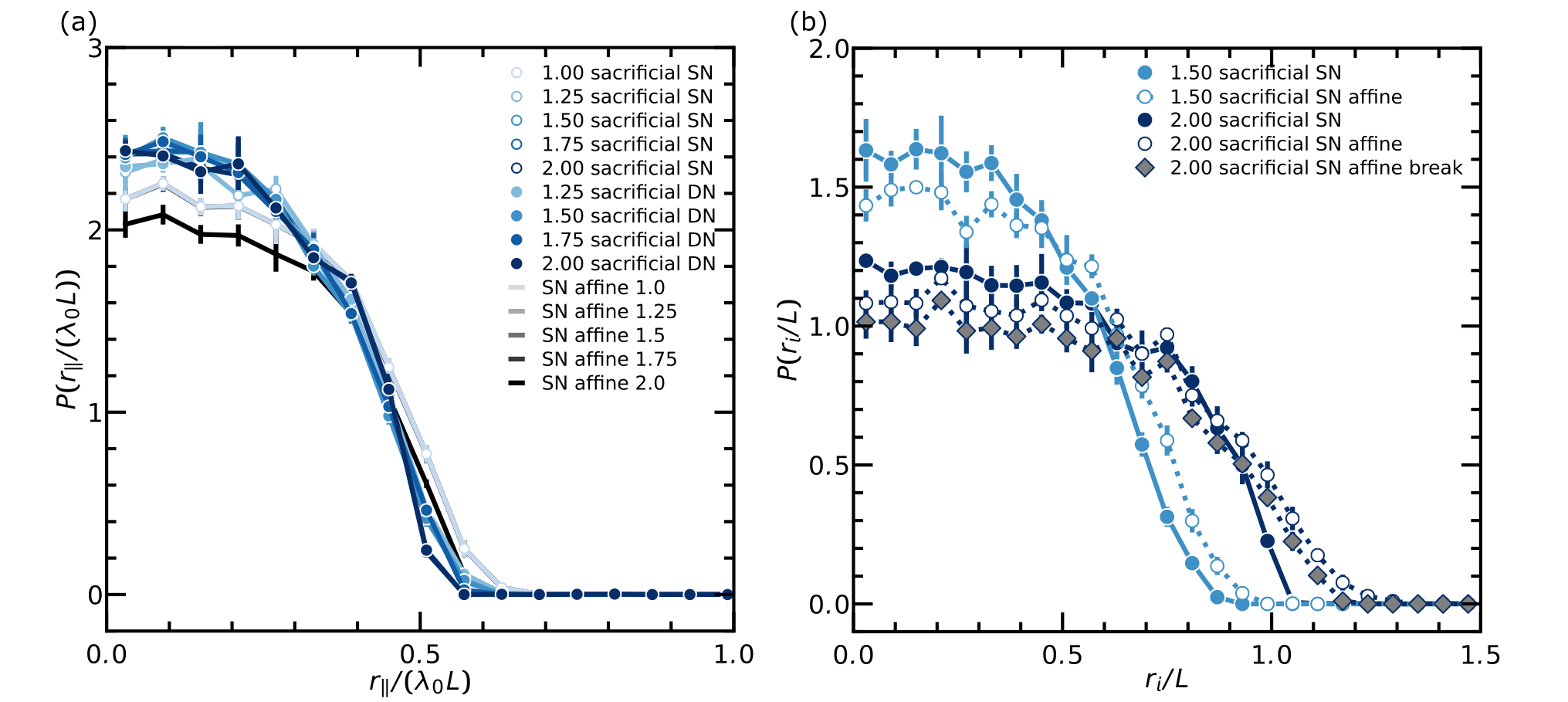}
    \caption{Evolution of pre-stretch during the swelling procedure. The response to swelling is largely affine. (a) Rescaling of the average chain-stretch distribution of sacrificial chains prior to deformation for SNs (open symbols) and DNs (solid symbols). In addition we show the rescaling of the affine prediction for the average chain-stretch distribution based on affine swelling with respect to the $\lambda_0=1.00$ configuration (grey lines). As expected almost all affine predictions collapse on the pre-stretch distribution at $\lambda_0=1.00$. Only the affine prediction for $\lambda_0=2.00$ does not collapse, due to rupture of polymer chains. (b) Comparison between the average stretch distribution $r_i/L$ at \SI{0}{\percent} strain obtained from simulation (closed faced circles) and from affine swelling with respect to the average positions at $\lambda_0=1.00$ (open-faced circles). For $\lambda_0=2.00$ we also show the distribution for affine swelling where polymers that exceed $r/L=1.129$ are considered broken (grey faced diamonds).}
    \label{fig:ch6figS8}
\end{figure}



\begin{figure}
    \centering
    \includegraphics{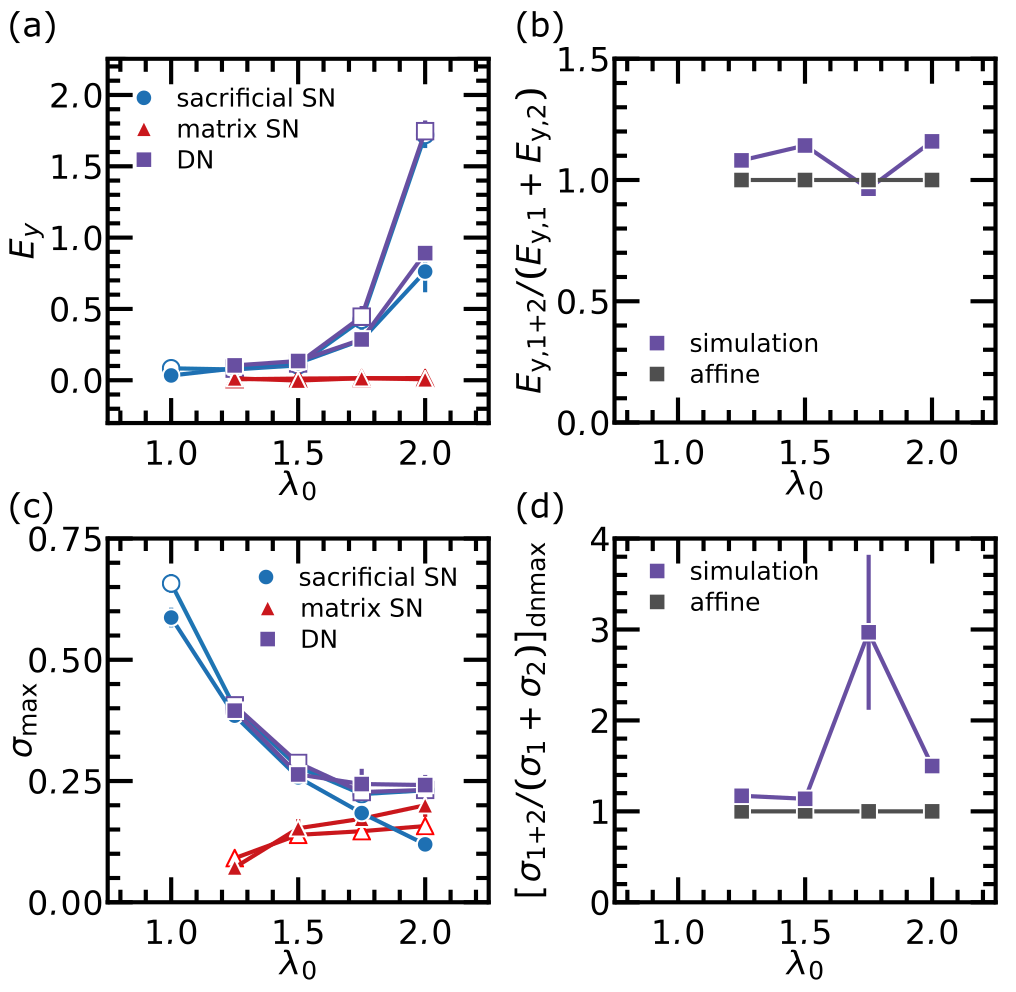}
    \caption{Enhancement in Young's modulus and peak strength as a function of $\lambda_0$. (a) Young’s modulus for sacrificial SN (blue), matrix SN (red), and DN (purple) versus the swelling ratio. Including simulation results (closed symbols) and affine predictions (open symbols). Error bars indicate the standard deviation over \num{4} configurations. To measure Young’s modulus in simulation, we apply a step strain protocol. The uniaxial strain is increased in steps ($\Delta\lambda=0.005$) at a strain rate $\dot{\epsilon} = \num{1e-4}$. Between the strain steps no deformation is applied and the system is equilibrated for $\num{10000}\tau$ (stress is reported every $\num{1}\tau$). At each strain step the average stress is calculated based on the final $\num{9500}\tau$ of each step. We calculate the modulus as the slope of a linear fit to the average stress at $\lambda=\{1.0,1.005,1.01\}$. For each configuration the modulus is calculated for deformation in the $x$, $y$ and $z$ direction and averaged. (b) Enhancement in $E_y$ of DNs with respect to SNs for simulations (purple) and affine prediction (grey). (c) Maximum stress $\sigma_{\text{max}}$ for sacrificial SN (blue), matrix SN (red), and DN (purple) versus the swelling ratio. Including simulation results (closed symbols) and affine predictions (open symbols). Errorbars indicate the standard deviation over \num{4} configurations. (d) Enhancement in $\sigma_{\text{max}}$ of DNs with respect to SNs for simulations (purple) and affine prediction (grey). For this panel the $\sigma$ values of the SNs are obtained at the strain at maximum stress of the DN.}
    \label{fig:ch6figS3}
\end{figure}


\begin{figure}
    \centering
    \includegraphics{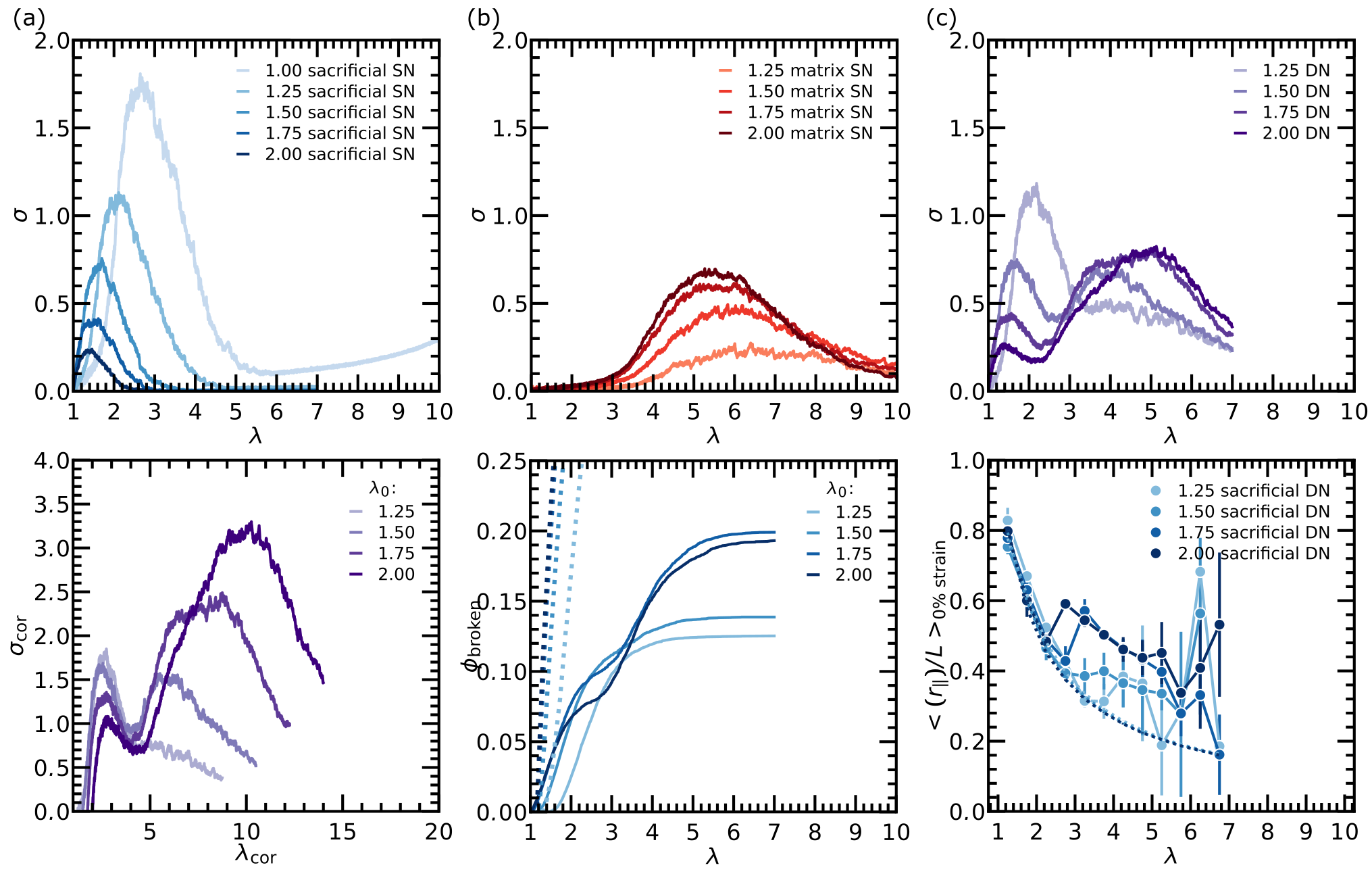}
    \caption{Mechanical response of networks formed at $\rho=0.34$ ($c_1=5\%$, $c_2=1\%$), instead of $\rho=0.15$ as shown in the main text. Qualitatively similar conclusions can be drawn also in this case.  However, we noticed that higher density networks at $c_1=\SI{5}{\percent}$ accumulate a significant amount of damage if they are swollen beyond $\lambda_0=1.25$. Data are averaged over \num{4} configurations. For every configuration simulations are performed by uniaxial extension in the $x$, $y$ and $z$ direction and the response is averaged. (a) Stress-strain response for the sacrificial SNs. (b) Stress-strain response for the matrix SNs. (c) Stress-strain response for the DNs. (d) Rescaled DN data. (e) fraction of broken sacrificial chains in simulations on DNs and corresponding the affine prediction. (f) Average initial pre-stretch of broken sacrificial chains as a function of the global strain for simulations on DNs.}
    \label{fig:ch6figS4}
\end{figure}

\begin{figure}
    \centering
    \includegraphics{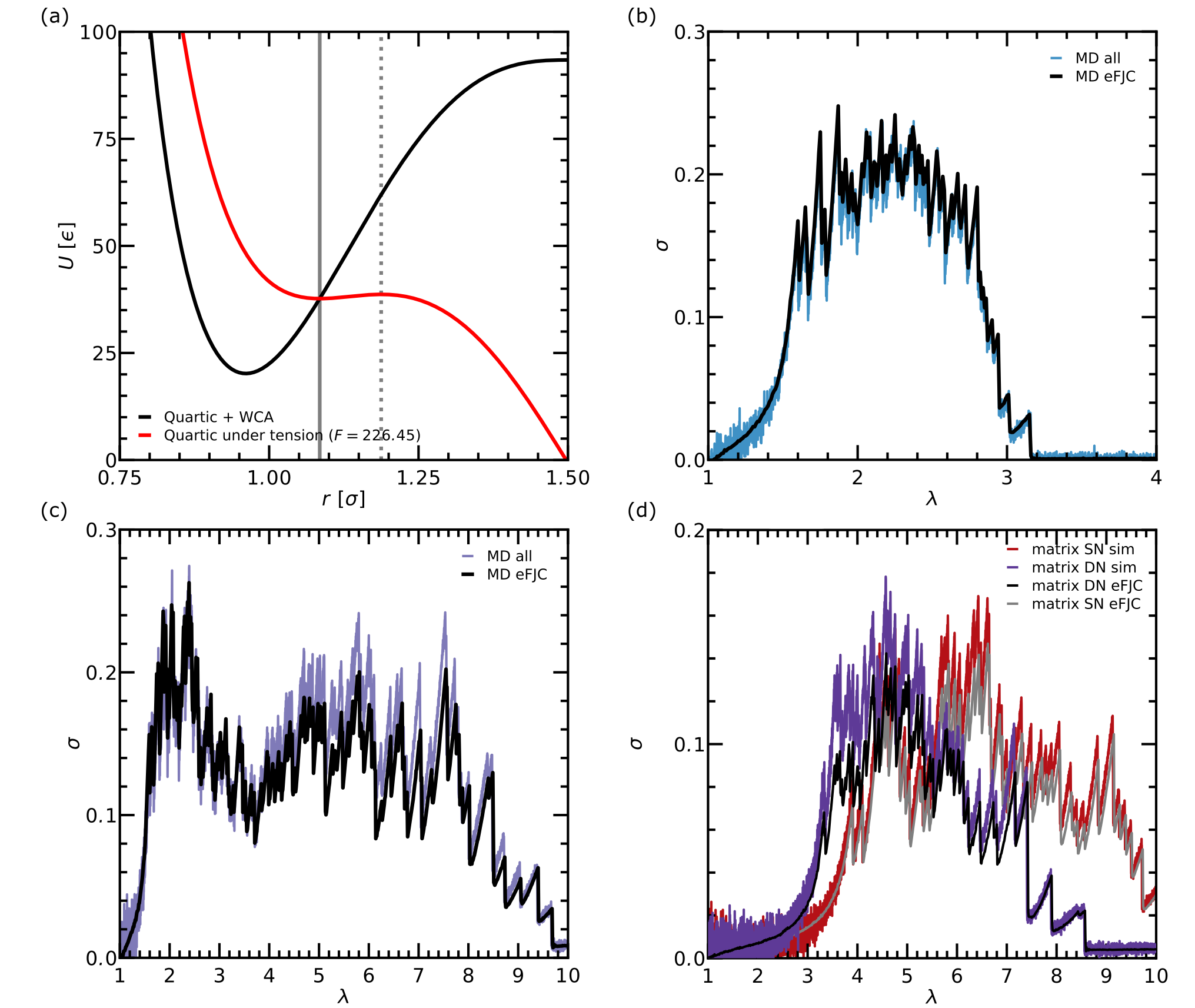}
    \caption{Estimation of the stress-strain response by mapping the end-to-end distances on extensible freely jointed chains (eFJC). Although our short chains are not expected to behave exactly as ideal chains, our eFJC estimation agrees well with the simulation results.  (a) Quartic potential used in our simulations (black), quartic potential with coupled work due to tension $U_{\text{tension}}(R) = U(R) - F\times(R-R_{R})$ (red). With $R_R$ the rest length under tension (solid grey line). At $F=226.45$ the energy barrier to bond rupture is $1.00$ (in reduced units). (b) and (c) Comparison of the stress-strain response in simulation (blue) and the eFJC estimation (black). Data is shown for a single configuration and deformation along the $x$-axis. (b) SN response ($\lambda_0=1.50$, $\rho=0.15$, $c_1=5\%$) (c) DN response ($\lambda_0=1.50$, $\rho=0.15$, $c_1=5\%$, $c_2=1\%$). (d) For the DN the eFJC approach (slightly) underestimates the stress response, because the end-to-end distance underestimates the tension in chains that are affected by topological constraint. This effect is most prominent for matrix chains, which we show here for a network at $\lambda_0=1.75$ ($\rho=0.15$, $c_1=5\%$, $c_2=1\%$). If we only plot the contribution of the matrix chains to the global stress in a DN, we clearly see that the eFJC estimation (black) underestimates the stress response (red). We also observe that in the matrix SN the difference between the stress response (red) and the eFJC estimation (grey) is smaller, suggesting that the difference in DNs is mainly caused by inter-network constraints.}
    \label{fig:ch6figS6}
\end{figure}


\begin{figure}
    \centering
    \includegraphics{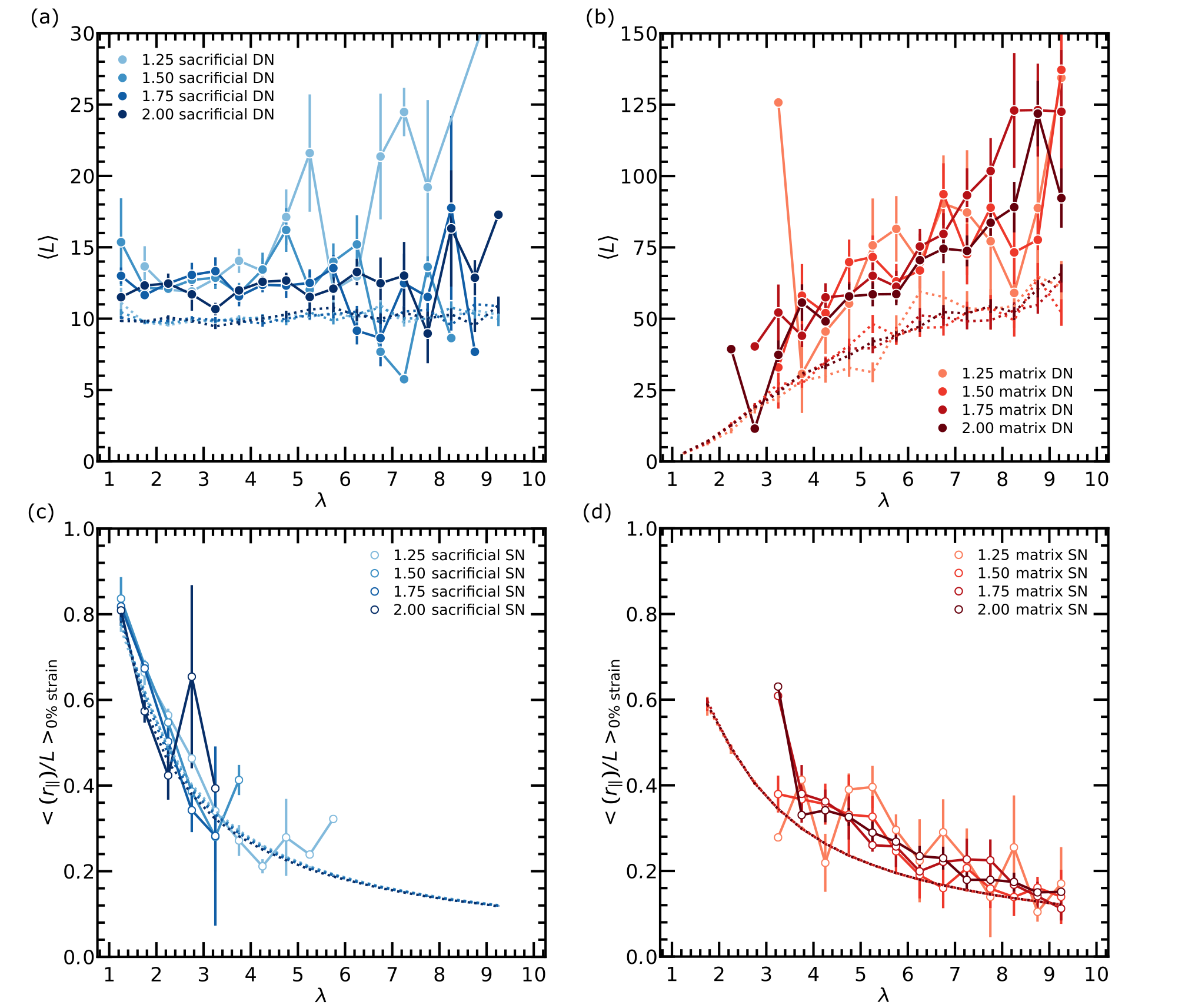}
    \caption{(a) and (b) Average length of broken chains as a function of global strain found in simulation (solid lines) and according to the affine prediction (dashed lines). (a) Sacrificial chains in DN networks. (b) Matrix chains in DN networks. (c) and (d)  Average initial pre-stretch of broken chains as a function of the global strain at break found in simulation (solid lines) and according to the affine prediction (dashed lines). (c) Sacrificial chains in SN networks. (d) Matrix chains in SN networks.}
    \label{fig:ch6figS7}
\end{figure}

\begin{figure}
    \centering
    \includegraphics{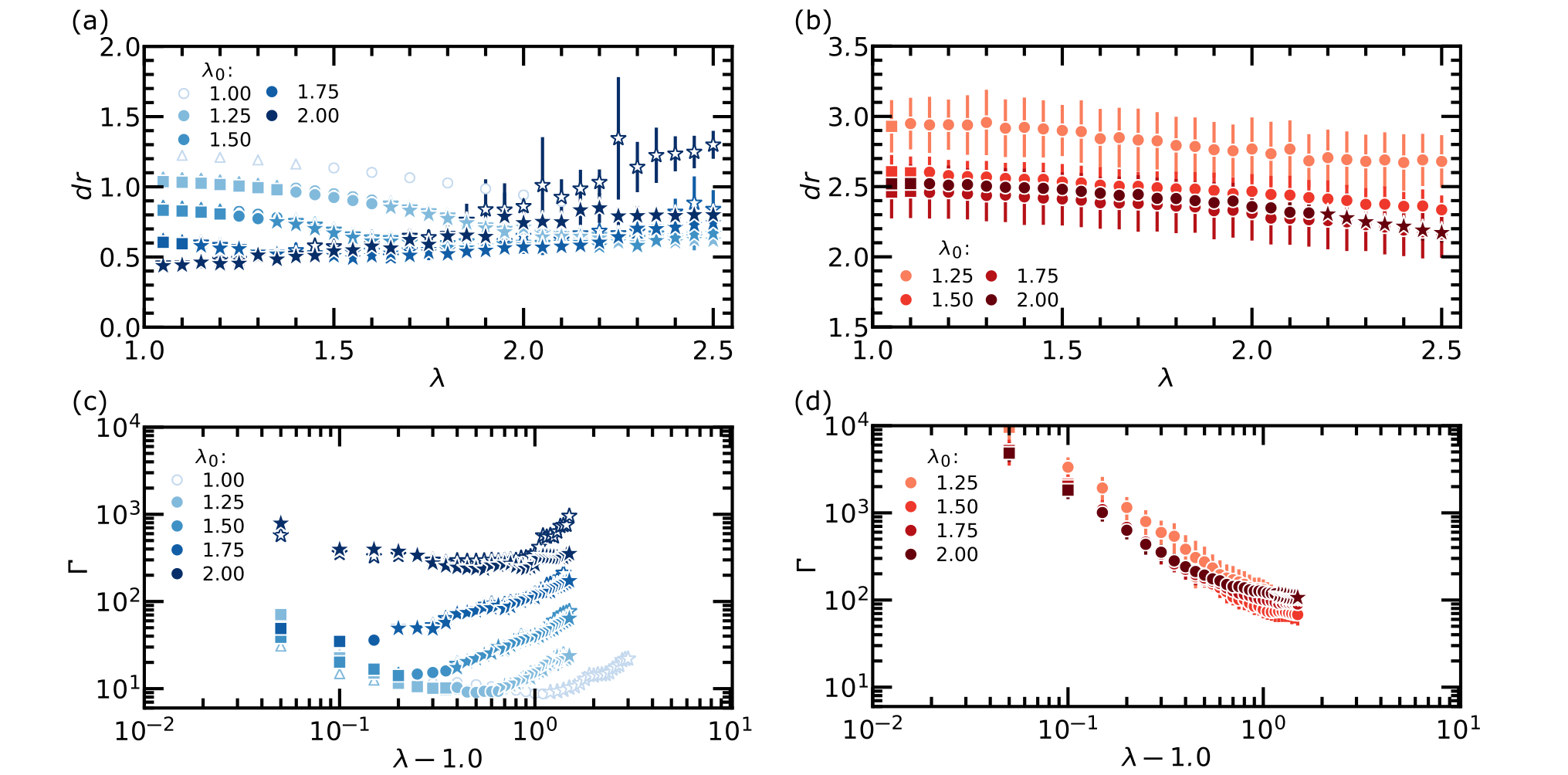}
    \caption{Effect of $\lambda_0$ and uniaxial strain on thermal and non-affine fluctuations of crosslinkers. With $\lambda_0$ the size of the thermal fluctuation in the sacrificial network decreases, while the size of the non-affine rearrangements increases. A large part of the non-affine response is either dominated by thermal fluctuations or by rupture of polymer chains. We determine the size of these fluctuations from the crosslinker positions obtained during a step-strain protocol where every strain step ($\Delta\lambda=0.05$) the system is equilibrated for $\num{10000}\tau$ and crosslinker positions are reported every $\num{50}\tau$. Only crosslinkers that are part of the largest percolating cluster are considered. The particle positions in each snapshot are unwrapped and corrected for the center of mass (only considering the crosslinkers). Using these coordinates we can calculate the average position of every crosslinker $\langle \mathbf{r} \rangle$ and the size of the thermal fluctuations around this position $dr = \langle \sqrt{(\mathbf{r}-\langle\mathbf{r}\rangle)^2}\rangle$ as a function of strain. In the following plots the marker shape indicates different regimes: around squares the non-affine fluctuations are smaller than the thermal fluctuations, circles represent the elastic regime, and around stars breaking of chains has started. (a) Size of the thermal fluctuations in the sacrificial SNs (open symbols and sacrificial DN (closed symbols) (b) Size of the thermal fluctuations in matrix DNs. (c) and (d) Non-affinity parameter $\Gamma = \langle (\mathbf{u} - \mathbf{u}_{\text{aff}})^2 \rangle / (\epsilon^2 \sigma^2)$. With $u$ the displacement of the average crosslinker position $\langle \mathbf{r} \rangle$ with respect to $\lambda=1.00$ and $u_{\text{aff}}$ the displacement under affine conditions.  (c) $\Gamma$ for the sacrificial SNs (open symbols) and sacrificial DN (closed symbols) (d) $\Gamma$ for matrix DNs. }
    \label{fig:ch6figS9}
\end{figure}

\clearpage
\bibliography{supp}